\documentclass[aip,jcp,preprint,superscriptaddress,noshowkeys]{revtex4-2}

\usepackage{graphicx,dcolumn,bm,xcolor,microtype,multirow,amscd,amsmath,amssymb,amsfonts,physics,wrapfig,bbold,siunitx,xspace}
\usepackage[version=4]{mhchem}

\usepackage[utf8]{inputenc}
\usepackage[T1]{fontenc}

\usepackage{hyperref}
\hypersetup{
    colorlinks,
    linkcolor={red!50!black},
    citecolor={red!70!black},
    urlcolor={red!80!black}
}

\usepackage{listings}
\definecolor{codegreen}{rgb}{0.58,0.4,0.2}
\definecolor{codegray}{rgb}{0.5,0.5,0.5}
\definecolor{codepurple}{rgb}{0.25,0.35,0.55}
\definecolor{codeblue}{rgb}{0.30,0.60,0.8}
\definecolor{backcolour}{rgb}{0.98,0.98,0.98}
\definecolor{mygray}{rgb}{0.5,0.5,0.5}

\definecolor{sqred}{rgb}{0.85,0.1,0.1}
\definecolor{sqgreen}{rgb}{0.25,0.65,0.15}
\definecolor{sqorange}{rgb}{0.90,0.50,0.15}
\definecolor{sqblue}{rgb}{0.10,0.3,0.60}

\lstdefinestyle{mystyle}{
    backgroundcolor=\color{backcolour},
    commentstyle=\color{codegreen},
    keywordstyle=\color{codeblue},
    numberstyle=\tiny\color{codegray},
    stringstyle=\color{codepurple},
    basicstyle=\ttfamily\footnotesize,
    breakatwhitespace=false,
    breaklines=true,
    captionpos=b,
    keepspaces=true,
    numbers=left,
    numbersep=5pt,
    numberstyle=\ttfamily\tiny\color{mygray},
    showspaces=false,
    showstringspaces=false,
    showtabs=false,
    tabsize=2
  }

  \newcolumntype{d}{D{.}{.}{-1}}

\usepackage{booktabs}
\usepackage{array}
\usepackage{multirow}

\newcolumntype{C}[1]{>{\centering\arraybackslash}p{#1}}  
\newcolumntype{R}[1]{>{\raggedleft\arraybackslash}p{#1}} 

\usepackage{physics}

\newcommand{\hH}{\Hat{H}}
\newcommand{\hZ}{\Hat{Z}}
\newcommand{\stH}{\Bar{H}}
\newcommand{\hT}{\Hat{T}}
\newcommand{\cre}[1]{\Hat{a}^{\dagger}_{#1}}
\newcommand{\ani}[1]{\Hat{a}_{#1}}
\newcommand{\Sp}[1]{\Hat{P}^{+}_{#1}}
\newcommand{\Sm}[1]{\Hat{P}^{-}_{#1}}
\newcommand{\Spm}[1]{\Hat{P}^{\pm}_{#1}}
\newcommand{\n}[1]{\Hat{n}_{#1}}
\newcommand{\SD}{\Phi} 
\newcommand{\PP}{\Psi} 
\newcommand{\pCCD}{\Phi_\text{pCCD}} 
\newcommand{\ger}[1]{#1_{0}}
\newcommand{\ung}[1]{#1_{1}}

\begin{document}
	
	\title{Connections between Richardson-Gaudin States, Perfect-Pairing, and Pair Coupled-Cluster Theory}
	\author{Paul A. Johnson}
	\email{paul.johnson@chm.ulaval.ca}
	\author{Charles-\'{E}mile Fecteau}
	\author{Samuel Nadeau}
	\affiliation{D\'{e}partement de chimie, Universit\'{e} Laval, Qu\'{e}bec, Qu\'{e}bec, G1V 0A6, Canada}
	
	\author{Mauricio Rodr\'{i}guez-Mayorga}
	
	\author{Pierre-Fran\c{c}ois Loos}
	\affiliation{Laboratoire de Chimie et Physique Quantiques (UMR 5626), Universit\'{e} de Toulouse, CNRS, Toulouse, France}
	
	\begin{abstract}
	Slater determinants underpin most electronic structure methods, but orbital-based approaches often struggle to describe strong correlation efficiently.
	Geminal-based theories, by contrast, naturally capture static correlation in bond-breaking and multireference problems, though at the expense of implementation complexity and limited treatment of dynamic effects.
	In this work, we examine the interplay between orbital and geminal frameworks, focusing on perfect-pairing (PP) wavefunctions and their relation to pair coupled-cluster doubles (pCCD) and Richardson-Gaudin (RG) states.
	We show that PP arises as an eigenvector of a simplified reduced Bardeen-Cooper-Schrieffer (BCS) Hamiltonian expressed in bonding/antibonding orbital pairs, with the complementary eigenvectors enabling a systematic treatment of weak correlation.
	Second-order Epstein-Nesbet perturbation theory on top of PP is found to yield energies nearly equivalent to pCCD.
	These results clarify the role of pair-based ans\"atze and open avenues for hybrid approaches that combine the strengths of orbital- and geminal-based methods.	\end{abstract}
	
	\maketitle
	
	\section{Introduction}

	Slater determinants, built from single-electron orbital wavefunctions, play a central role in quantum chemistry as the building blocks for most correlation treatments. \cite{helgaker_book}
	These include perturbation theory, configuration interaction (CI), multiconfigurational self-consistent field (MCSCF), and coupled-cluster (CC) theory. \cite{helgaker_book,bartlett_book}
	Their widespread use is largely due to the efficiency of second quantization, \cite{dirac:1927,jorgersen_book} which, along with practical tools such as Wick's theorem, \cite{wick:1950,crawford:2000}
	facilitates theoretical developments and efficient computational implementations in electronic structure methods. \cite{helgaker_book,hirata:2003,rubin:2011,evangelista:2022,quintero-monsebaiz:2023,liebenthal:2025}

	Despite their theoretical appeal, these approaches, while systematically improvable, often struggle to fully capture electron correlation in practice. 
	Low-order methods such as MP2, CISD, and CCSD describe weak (dynamic) correlation well but fail to account for strong (static) correlation effects. 
	Accurately treating strong correlation requires more advanced methods, such as complete active space self-consistent field (CASSCF), \cite{roos:1980a,roos:1980b,roos:2005} density matrix renormalization group (DMRG), \cite{white:1992,white:1993,chan:2011,wouters:2014,szalay:2015,baiardi:2020} higher-order CC methods (e.g., CCSDTQ), \cite{oliphant:1991,kucharski:1992} or selected CI approaches. \cite{huron:1973,giner:2013,liu:2014,schriber:2016,holmes:2016,garniron:2018}
	However, the computational cost of these methods limits their application to relatively small systems.

	An alternative perspective comes from geminal-based theories, \cite{hurley:1953,shull:1959,allen:1961,coleman:1963} which explicitly model electron pairing and offer a conceptually appealing way to describe strong correlation. 
	Because of their direct connection to chemical bonding and Lewis structures, \cite{lewis:1916} geminal-based methods naturally capture strong correlation in bond-breaking and multi-reference problems. \cite{surjan_book,coleman:1965,silver:1969,limacher:2013,tecmer:2014,johnson:2020,debaerdemacker:2024}
	However, their implementation is challenging due to the complexity of handling geminal wavefunctions, and they often neglect weak correlation effects. 
	As a result, electronic structure theory remains largely dominated by orbital- and density-based approaches.
	
	Bridging the gap between orbital and geminal methods remains a key challenge in quantum chemistry. 
	A deeper understanding of their connections may help develop hybrid approaches that combine the advantages of both, i.e., retaining the systematic improvability of orbital-based methods while incorporating the robust treatment of strong correlation inherent in geminal theories. \cite{cullen:1996,thorsteinsson:1996a,thorsteinsson:1996b,thorsteinsson:1997,larsson:2020,dunning:2016}
	
	The current push for pair theories stems from Ref.~\onlinecite{bytautas:2011}, which demonstrated that strongly-correlated systems could be described efficiently in terms of CI expansions in which the Slater determinants were classified in terms of their number of unpaired electrons, their \emph{seniority}, rather than their level of excitation from a given reference. Seniority-zero CI or doubly-occupied CI (DOCI), i.e., a CI of Slater determinants with no unpaired electrons (or closed-shell determinants), \cite{allen:1961,smith:1965,veillard:1967} was found to describe bond-breaking processes qualitatively with systematic improvement possible by adding seniorities two and four.\cite{bytautas:2015,alcoba:2014a,alcoba:2014b,kossoski:2022,kossoski:2023} Unfortunately, even seniority-zero CI has exponential cost, 
	though a cheap method was quickly developed with almost no loss in numerical accuracy. This was published as the antisymmetric product of 1-reference orbital geminals (AP1roG), \cite{limacher:2013} and was immediately understood to be equivalent to pair coupled-cluster doubles (pCCD). \cite{stein:2014,henderson:2014b} 
	Seniority-zero methods, hence pCCD, are not invariant to orbital rotations and thus require an orbital optimization.
	\cite{boguslawski:2014a,boguslawski:2014b,boguslawski:2014c,henderson:2014b,boguslawski:2016a,kossoski:2021,marie:2021}
	As a result, the reference is \emph{not} the Hartree-Fock (HF) Slater determinant. The optimal orbitals are generally localized on at most two atomic centers, which resemble those from generalized valence bond (GVB) theory.\cite{limacher:2014a}
	As defined by Dunning and coworkers, \cite{dunning:2016} GVB is a linear combination (i.e., a CI expansion) of all possible singlet-coupled configuration state functions (CSFs) in a given set of orbitals. This definition necessarily includes contributions from all possible seniorities.
	Usually, only a small number of CSFs are relevant, which leads to the practical approximation of \emph{strongly-orthogonal perfect pairs}. In this case, GVB simplifies to a \emph{product} of singlet-coupled electron pairs: each coupling contains two spatial orbitals, while each spatial orbital only appears in one such coupling. 
	In terms of atomic orbitals, this state is a product of Heitler-London-type states, while its natural orbitals are localized into sets of bonding and antibonding orbitals. Henceforth, we will only consider GVB with strongly-orthogonal perfect pairs and refer to it as perfect-pairing (PP).	
	PP is understood to describe strong correlation in bond-breaking processes.\cite{hurley:1953,goddard:1967,hunt:1972,hay:1972,goddard:1973,bobrowicz:1977,goddard:1978,dykstra:1980,small:2009,lawler:2010,small:2011,lehtola:2016,beran:2005,beran:2006,parkhill:2009,parkhill:2010a,parkhill:2010b,dunning:2016} 
	It has long been recognized that PP can be written as a CC wavefunction. \cite{cullen:1996,cullen:1999,vanvoorhis:2000,vanvoorhis:2000b,vanvoorhis:2001,small:2012,cullen:2007} 
	More recently, PP-based ans\"atze have proven useful to maximize the accuracy of shallow quantum circuits in the context of quantum computing. \cite{kottmann:2022,kottmann:2023,kottmann:2024,burton:2024a,burton:2024b}
	
	In a previous paper, \cite{johnson:2024b} it was demonstrated that DOCI can be reduced to single-reference Epstein-Nesbet perturbation theory (ENPT) \cite{epstein:1926,nesbet:1955} built from eigenvectors of the reduced Bardeen-Cooper-Schrieffer (BCS) Hamiltonian,\cite{cooper:1956,bardeen:1957a,bardeen:1957b} the so-called Richardson-Gaudin (RG) states.\cite{richardson:1963,richardson:1964,richardson:1965,gaudin:1976} In this manuscript, we will first show that the reduced BCS Hamiltonian can be simplified to a form describing bonding/antibonding pairs of orbitals. PP emerges as an eigenvector of the simplified model, while the other eigenvectors provide a basis for PP's orthogonal complement.	Weak correlation, in the PP picture, can then be accounted for systematically in a single-reference approach. In particular, it will be demonstrated that the corresponding second-order ENPT (EN2) PP corrected energy is near-equivalent to its pCCD counterpart.

	\section{Theory}
	\subsection{Perfect-Pairing}
	
	First, we will fix our notation. We consider closed-shell systems composed of $N$ electrons. Further, we assume there is no core so that for a system with $K$ spatial orbitals, the $N$ electrons are distributed in $N$ active spatial orbitals composed of bonding and antibonding pairs (see below), while the $K-N$ remaining spatial orbitals are labeled as virtuals.
	The indices $1 \le p,q,r,s,\ldots \le N$ indicate active spatial orbitals, while the sets of indices $1 \le i,j,k,l,\ldots \le N/2$ and $N/2 + 1 \le a,b,c,d,\ldots \le N$ denote occupied and unoccupied active spatial orbitals, respectively.
	The Greek letters $1 \le \alpha,\beta,\gamma,\delta,\ldots \le N/2$ are reserved for orbital pairs and $\sigma,\tau =$ $\uparrow$ or $\downarrow$ are spin labels.
	The letters $\mu,\nu,\lambda$ will only refer to the bonding/antibonding labels, taking values either 0 or 1.
	
	Computations involving Slater determinants --- many-electron wavefunctions written as antisymmetrized sums of products of one-electron orbitals --- can be conveniently reduced to operations with creation (annihilation) operators $\cre{p\sigma}$ ($\ani{p\sigma}$) which add (remove) an electron in spatial orbital $p$ with spin $\sigma$. 
	These one-body operators fulfill well-known anticommutation rules: $\{ \cre{p\sigma}, \cre{q\tau}\} = 0$, $\{ \ani{p\sigma}, \ani{q\tau}\} = 0$, and $\{ \cre{p\sigma}, \ani{q\tau}\} = \delta_{pq}\delta_{\sigma\tau}$, where $\{ \cdot, \cdot \}$ is the anticommutator.
	
\begin{figure*}
		\includegraphics[height=0.50\linewidth]{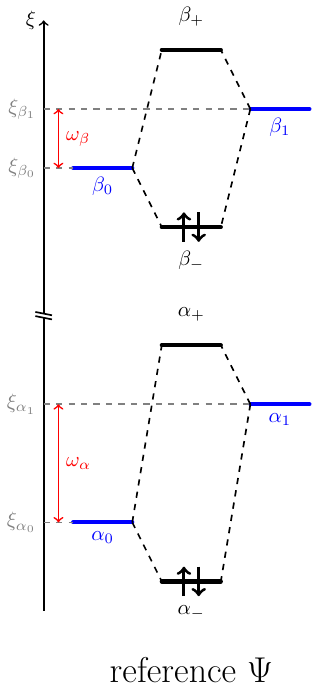}
		\vspace{0.05\linewidth}
		\\
		\includegraphics[height=0.50\linewidth]{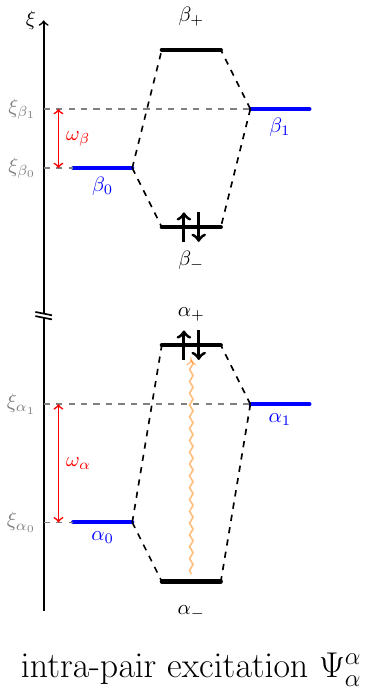}
		\hspace{0.05\linewidth}
		\includegraphics[height=0.50\linewidth]{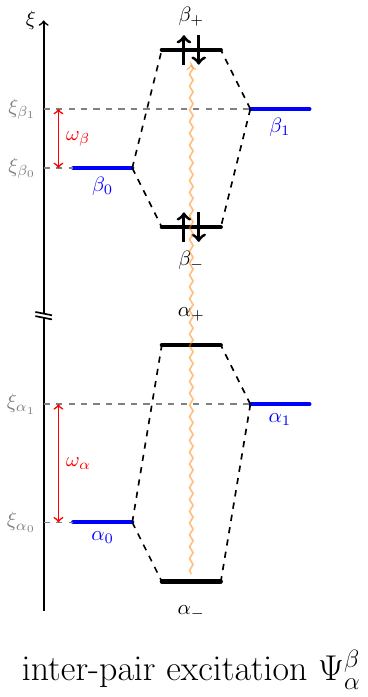}
		\hspace{0.05\linewidth}
		\includegraphics[height=0.50\linewidth]{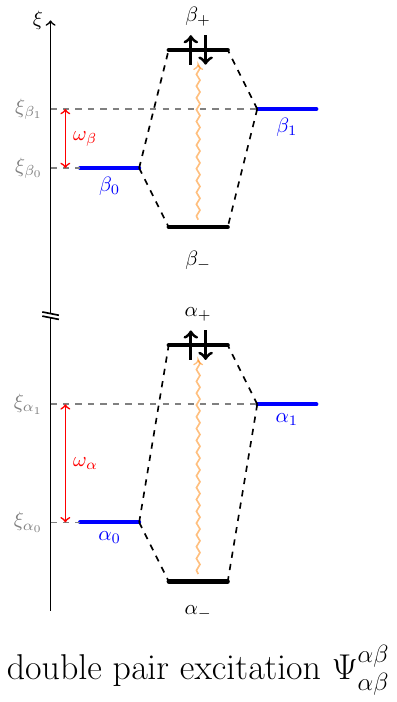}
		\caption{Schematic representation of the orbital diagram in PP: reference PP wave function (top) and the three types of excited configurations (bottom). Here, we have represented two VBS, $\{ \ger{\alpha}, \ung{\alpha} \}$ and $\{ \ger{\beta}, \ung{\beta} \}$, characterized by VBS gaps $\omega_\alpha = \xi_{\ung{\alpha}} - \xi_{\ger{\alpha}}$ and $\omega_\beta = \xi_{\ung{\beta}} - \xi_{\ger{\beta}}$, respectively.}
		\label{fig:sketch}
	\end{figure*}

	Computations with pair (i.e., two-electron) wavefunctions, also known as geminals, are more involved but can be reduced to operations with the three operators
	\begin{align} \label{eq:pair_su2}
		\Sp{p} & = \cre{p\uparrow} \cre{p\downarrow} 
		&
		\Sm{p} & = \ani{p\downarrow} \ani{p\uparrow}
		&
		\n{p} & = \cre{p\uparrow} \ani{p\uparrow} + \cre{p\downarrow}\ani{p\downarrow}
	\end{align}
	where $\Sp{p}$ ($\Sm{p}$) creates (removes) a pair of opposite-spin electrons in the spatial orbital $p$, while $\n{p}$ counts the number of electrons in the spatial orbital $p$. These three operators have the structure of the Lie algebra su(2)
	\begin{align}
		\comm{\Sp{p}}{\Sm{q}} &= \delta_{pq} \qty( \n{p} -1 ) 
		&
		\comm{\n{p}}{\Spm{q}} &= \pm 2 \delta_{pq} \Spm{p}.
	\end{align}	
	The reduced BCS Hamiltonian
	\begin{align} \label{eq:H_BCS}
		\hH_{\text{BCS}} = \frac{1}{2} \sum_{p} \xi_p \n{p} + \frac{1}{2} \sum_{pq} \Sp{p} \Sm{q}
	\end{align}
	is an exactly solvable model describing pairs of electrons distributed among spatial orbitals with energies $\{\xi_p\}$ in a constant strength interaction.	
	Its eigenvectors, the RG states, are simply constructed from non-linear equations. In previous papers,\cite{johnson:2020,moisset:2022a,fecteau:2022,johnson:2023,johnson:2024a,johnson:2024b} molecular systems were treated with RG states by variationally minimizing the energy of the Coulomb Hamiltonian
	\begin{equation} \label{eq:Hc}
		\hH_\text{C} = \sum_{pq} h_{pq} \sum_{\sigma} \cre{p\sigma} \ani{q\sigma} 
		+ \frac{1}{2} \sum_{pqrs} V_{pqrs} \sum_{\sigma \tau} \cre{p \sigma} \cre{r \tau} \ani{s \tau} \ani{q\sigma} 
	\end{equation}
	(where $h_{pq}$ are the elements of the one-electron Hamiltonian and $V_{pqrs}$ are direct two-electron integrals in Mulliken's, or chemists', notation) with respect to the parameters $\{\xi_p\}$. In other words, an optimal model Hamiltonian was identified that produced a state to minimize the energy of the real molecular system. 	
	The optimal $\{\xi_p\}$ found grouped the $N$ spatial orbitals into $N/2$ well-separated subsystems, each consisting of near-degenerate bonding and antibonding orbitals that we will call valence-bond subsystems (VBS).
	This is illustrated in Fig.~\ref{fig:sketch}.
	
	The reduced BCS Hamiltonian and its eigenvectors simplify dramatically when the VBS are treated as disjoint, leading to a collection of two-level systems that do not interact with one another. In other words, an \emph{independent-pair approximation}. Each of the $\alpha=1,\dots,N/2$ independent VBS contains a bonding spatial orbital $\ger{\alpha}$ and an antibonding spatial orbital $\ung{\alpha}$ (see Fig.~\ref{fig:sketch}). The number of pairs in each VBS is now a symmetry, and thus $\xi_{\alpha_0}(\n{\ger{\alpha}} + \n{\ung{\alpha}})$ is a constant that can be removed, in addition to the diagonal terms of the interaction. The effective Hamiltonian is thus 
	\begin{equation} \label{eq:H_GVB}
		\hH_\text{PP} = \frac{1}{2} \sum_{\alpha} \omega_\alpha \n{\ung{\alpha}} + \frac{1}{2} \sum_{\alpha} 
		\qty( \Sp{\ger{\alpha}} \Sm{\ung{\alpha}} + \Sp{\ung{\alpha}} \Sm{\ger{\alpha}} ).
	\end{equation}
	Pairs move between the bonding and antibonding orbitals with a constant strength, accounting for the Coulomb repulsion, while electrons in the antibonding orbital increase the energy by the VBS gap $\omega_\alpha = \xi_{\ung{\alpha}} - \xi_{\ger{\alpha}}$. As Eq.~\eqref{eq:H_GVB} is a sum of disjoint contributions for each VBS, its eigenvectors factor into products of the eigenvectors of each VBS. In the seniority-zero sector, i.e. in terms of only $\Sp{p}$, $\Sm{p}$ and $\n{p}$, each VBS has four eigenvectors: both orbitals can be empty $\ket{0}$, both orbitals can be doubly-occupied $\ket{\ger{\alpha}\ung{\alpha}} = \Sp{\ger{\alpha}} \Sp{\ung{\alpha}} \ket{0}$, or one pair can be distributed among the two orbitals, $\ket{\ger{\alpha}} = \Sp{\ger{\alpha}} \ket{0}$ and $\ket{\ung{\alpha}} = \Sp{\ung{\alpha}} \ket{0}$, in the two ways
	\begin{equation} \label{eq:ket_nua}
		\ket{\alpha_{\pm}} 
		= \ket{\ger{\alpha}} + \qty(\omega_\alpha \pm \sqrt{\omega^2_\alpha + 1}) \ket{\ung{\alpha}}
	\end{equation}
	based on the value of $\omega_\alpha$, with norm
	\begin{equation}
		\mathcal{N}_{\alpha_{\pm}} = \braket{\alpha_{\pm}}{\alpha_{\pm}} = 
		2 \sqrt{\omega^2_{\alpha} + 1} \qty(\sqrt{\omega^2_{\alpha} +1 } \pm \omega_{\alpha} ).
	\end{equation}
	The weights of $\ket{\ger{\alpha}}$ and $\ket{\ung{\alpha}}$ in Eq.~\eqref{eq:ket_nua} are chosen as, in the degenerate limit (i.e., $\omega_\alpha \to 0$), these two states become strict symmetric/antisymmetric linear combinations $(\ket{\ger{\alpha}} \pm \ket{\ung{\alpha}})/\sqrt{2}$. For $\ket{\alpha_{-}}$, the bonding orbital is more strongly occupied and the product
	\begin{equation} \label{eq:pp_def}
		\ket{\PP} = \prod_{\alpha} \ket{\alpha_{-}}
	\end{equation} 
	is the state traditionally called PP\footnote{The usual definition for PP is less precise: it is a product of pairs, each localized in two orbitals. When optimized for a repulsive interaction, this definition will always reduce to the state defined in Eq.~\eqref{eq:pp_def}} (see Fig.~\ref{fig:sketch}). It is an eigenvector of Eq.~\eqref{eq:H_GVB}, almost always the ground state. (If $\omega_\alpha \ll \omega_\beta$, the configuration $\ket{\ger{\alpha}\ung{\alpha}}$ is energetically more favorable than $\ket{\alpha_{-}\beta_{-}}$. For molecular systems, this does not happen, and $\ket{\PP}$ is the ground state of Eq.~\eqref{eq:H_GVB}.) 
	
	With $\mu \in \{0,1\}$, the one-electron reduced density matrix (1-RDM) elements are explicit functions of the VBS gaps
	\begin{equation} \label{eq:GVB_1dm}
		n_{\alpha_\mu} 
		= \frac{1}{2} \frac{\mel{\PP}{\n{\alpha_\mu}}{\PP}}{\braket{\PP}{\PP}} = \frac{1}{2} 
		\qty[ 1 + \frac{(-1)^\mu \omega_\alpha}{\sqrt{\omega^2_\alpha + 1}} ]
	\end{equation}
	(with $0 \le n_{\alpha_\mu} \le 1$ and $n_{\ger{\alpha}} + n_{\ung{\alpha}} = 1$) while there are two types of two-electron reduced density matrix (2-RDM) elements: the \emph{pair-transfer elements} are only non-zero within a VBS
	\begin{equation} \label{eq:GVB_2dm_P}
		P_{\ger{\alpha}\ung{\alpha}} 
		= \frac{\mel{\PP}{\Sp{\ger{\alpha}} \Sm{\ung{\alpha}}}{\PP}}{\braket{\PP}{\PP}} 
		= - \sqrt{n_{\ger{\alpha}} n_{\ung{\alpha}}}
		= - \frac{1}{2} \frac{1}{\sqrt{\omega^2_{\alpha} + 1}}
	\end{equation}
	and $P_{\alpha_\mu \alpha_\mu} = + n_{\alpha_\mu}$.
	The \emph{density-density elements} are only non-zero between distinct VBS (i.e., $\alpha \neq \beta$)
	\begin{equation} \label{eq:GVB_2dm_D}
		D_{\alpha_\mu \beta_\nu} 
		= \frac{1}{4} \frac{\mel{\PP}{\n{\alpha_\mu} \n{\beta_\nu}}{\PP}}{\braket{\PP}{\PP}} 
		= n_{\alpha_\mu} n_{\beta_\nu}.
	\end{equation}
	As $P_{\alpha_\mu \alpha_\nu}$ is only non-zero within a VBS and $D_{\alpha_{\mu}\beta_{\nu}}$ is reducible, there is no correlation between electron pairs for PP: the 2-RDM elements are explicit functions of the 1-RDM elements.\cite{piris:2011,pernal:2013} As the 1-RDM elements \eqref{eq:GVB_1dm} are themselves explicit functions of $\{\omega_\alpha\}$, the PP energy of the Coulomb Hamiltonian \eqref{eq:Hc}
	\begin{equation} \label{eq:E_GVB}
	\begin{split}
		E[\{\omega_\alpha\}] 
		& = 2 \sum_{\alpha} \sum_\mu n_{\alpha_\mu} \qty(h_{\alpha_\mu \alpha_\mu} + \frac{1}{2} L_{\alpha_\mu \alpha_\mu} )
		\\
		&+ 2 \sum_\alpha P_{\ger{\alpha}\ung{\alpha}} L_{\ger{\alpha} \ung{\alpha}}
		\\
		&+ 2 \sum_{\alpha<\beta} \sum_{\mu \nu} D_{\alpha_\mu \beta_\nu} G_{\alpha_\mu \beta_\nu}
	\end{split}
	\end{equation}
	can be minimized variationally with respect to the VBS gaps $\{\omega_\alpha\}$. Here, the two-electron integrals reduce to direct ($J$), exchange ($K$), and pair-transfer ($L$) types:
	\begin{subequations}
	\begin{align}
		J_{\alpha_\mu \beta_\nu} & = V_{\alpha_\mu \alpha_\mu \beta_\nu \beta_\nu}
		\\
		K_{\alpha_\mu \beta_\nu} & = V_{\alpha_\mu \beta_\nu \beta_\nu \alpha_\mu}
		\\
		L_{\alpha_\mu \beta_\nu} & = V_{\alpha_\mu \beta_\nu \alpha_\mu \beta_\nu}.
	\end{align}
	\end{subequations}
	For real orbitals, the exchange and pair-transfer integrals are identical. The three summations in Eq.~\eqref{eq:E_GVB} represent the one-body terms, the pair-transfer elements and the density-density elements, respectively.
	The direct and exchange integrals always occur together, so we adopt the shorthand
	\begin{align}
		G_{\alpha_\mu \beta_\nu} = 2J_{\alpha_\mu \beta_\nu} - K_{\alpha_\mu \beta_\nu}.	
	\end{align}
	We choose to write the diagonal two-electron integral $L_{\alpha_\mu \alpha_\mu} = V_{\alpha_\mu \alpha_\mu \alpha_\mu \alpha_\mu} $.

	To perform such a minimization, we rely on the Newton-Raphson algorithm, which requires the first (gradient)
	\begin{equation}  \label{eq:el_grad}
	\begin{split}
		\qty( \omega^2_\alpha + 1 )^{\frac{3}{2}} \pdv{E}{\omega_\alpha} 
		& = \omega_\alpha L_{\ger{\alpha} \ung{\alpha}} 
		+ \sum_\mu (-1)^\mu 
		\qty(
		h_{\alpha_\mu \alpha_\mu} + \frac{1}{2} L_{\alpha_\mu \alpha_\mu}+ \sum_{\beta (\neq \alpha)} \sum_{\nu} G_{\alpha_\mu \beta_\nu} n_{\beta_\nu}
		)
	\end{split}
	\end{equation}
	and second (Hessian) derivatives of the energy with respect to the VBS gaps
	\begin{subequations} \label{eq:el_hess}
	\begin{align} 
		& \qty( \omega^2_\alpha + 1 )^{\frac{3}{2}} \qty( \omega^2_\beta + 1 )^{\frac{3}{2}}
	 	\pdv[2]{E}{\omega_\alpha}{\omega_\beta} =
	 	\frac{1}{2} \sum_{\mu\nu} (-1)^{\mu+\nu} G_{\alpha_\mu \beta_\nu} \\
	 	& \qty( \omega^2_\alpha + 1 )^{\frac{3}{2}} \pdv[2]{E}{\omega_\alpha} =
	 	- 3 \omega_\alpha \qty( \omega^2_\alpha + 1 )^{\frac{1}{2}} \pdv{E}{\omega_\alpha}
	 	+ L_{\ger{\alpha} \ung{\alpha}}.
	 	\end{align}
	\end{subequations}
	Thus, for a given set of orbitals, the electronic gradient [see Eq.~\eqref{eq:el_grad}] and the electronic Hessian [see Eq.~\eqref{eq:el_hess}] are constructed with $\order{N^2}$ operations, which are cheaper than the linear algebra operations required for the Newton-Raphson steps.

	Requiring the gradient to vanish implies
	\begin{equation} \label{eq:grad_vanish}
		\sum_\mu (-1)^\mu 
		\qty(
		h_{\alpha_\mu \alpha_\mu} + \frac{1}{2} L_{\alpha_\mu \alpha_\mu}+ \sum_{\beta (\neq \alpha)} \sum_{\nu} G_{\alpha_\mu \beta_\nu} n_{\beta_\nu}
		)
		= - \omega_{\alpha} L_{\ger{\alpha} \ung{\alpha}},
	\end{equation}
	a property that will have consequences in perturbation theory (see below). Equation \eqref{eq:grad_vanish} gives a physical meaning to the VBS gaps: the left-hand side of Eq.~\eqref{eq:grad_vanish} is a difference of \emph{orbital energies} defined as
	\begin{equation}\label{eq:pp_orb_energies}
		\varepsilon_{\alpha_\mu} = h_{\alpha_\mu \alpha_\mu} + \frac{1}{2} L_{\alpha_\mu \alpha_\mu}+ \sum_{\beta (\neq \alpha)} \sum_{\nu} G_{\alpha_\mu \beta_\nu} n_{\beta_\nu}
	\end{equation}
	while $L_{\ger{\alpha} \ung{\alpha}}$ in the right-hand side of Eq.~\eqref{eq:grad_vanish} is the energy pushing a pair from the bonding orbital to the antibonding orbital. Here the orbital energies are distinct from those obtained from HF as they include occupation numbers between 0 and 1. Orbital energies have likewise been defined in various ways for pCCD.\cite{jahani:2025}
	
	The stationary VBS gaps are thus ratios of the differences in orbital energies to the Coulomb repulsion,
	\begin{equation}
		\omega_{\alpha} = \frac{\varepsilon_{\ung{\alpha}} - \varepsilon_{\ger{\alpha}}}{L_{\ger{\alpha} \ung{\alpha}}}
	\end{equation}
	just as they are in the model Hamiltonian \eqref{eq:H_GVB}. The stationary PP energy can then be recast
	\begin{equation} \label{eq:gvb_stat}
		E = \sum_{\alpha} \qty[
		\varepsilon_{\ger{\alpha}} + \varepsilon_{\ung{\alpha}} + (\varepsilon_{\ger{\alpha}} - \varepsilon_{\ung{\alpha}})
		\sqrt{1 + \qty(\frac{L_{\ger{\alpha} \ung{\alpha}}}{\varepsilon_{\ung{\alpha}} - \varepsilon_{\ger{\alpha}}})^2}
		]
		- 2 \sum_{\alpha < \beta} \sum_{\mu\nu} n_{\alpha_\mu} n_{\beta_\nu} G_{\alpha_\mu \beta_\nu},
	\end{equation}
	similar to an expression obtained by Kutzelnigg\cite{kutzelnigg:2012} for a two-level problem in particular. Equation \eqref{eq:gvb_stat} reduces to the stationary HF energy when the VBS gaps $\{\omega_{\alpha}\}$ become large. 
	\footnote{First, the bracketed term under the square root will vanish, leading to a root of 1, and the result of the first summation is 2 times the sum of the orbital energies of the bonding orbitals. In the second term, the occupation numbers will be the integers 1 (bonding) and 0 (anti-bonding) so that only the bonding orbitals contribute. In this limit, the PP state becomes a Slater determinant of fully occupied bonding orbitals.}
		
	Unfortunately, orbital optimization is necessary for pair wavefunctions. \cite{boguslawski:2014a,boguslawski:2014b,boguslawski:2014c,henderson:2014b,lehtola:2018} One must also compute the first and second derivatives of the energy with respect to the orbital rotation parameters $\{ \kappa_{pq} \}$. A unitary transformation of the orbitals 
	\begin{equation}
		\hat{U} = \exp 
		[
		\sum_{p<q} \kappa_{pq} \sum_{\sigma}
		\qty(
		\cre{p\sigma}\ani{q\sigma} - \cre{q\sigma}\ani{p\sigma}
		)
		]
	\end{equation}
	may be constructed as a step away from the identity $\boldsymbol{\kappa}=0$. In this case, the gradient and Hessian simplify to expectation values of commutators.\cite{helgaker_book} The orbital gradient
	\begin{equation} \label{eq:orb_grad}
	\begin{split} 
		\pdv{E}{\kappa_{pq}} 
		& = 4 h_{pq} (n_p - n_q) 
		\\
		& + 4 \sum_r (2V_{rrpq} - V_{rqpr})(D_{rp} - D_{rq})
		\\
		& + 4 \sum_r V_{rprq} (P_{rp} - P_{rq})
	\end{split}
	\end{equation}
	and Hessian
	\begin{equation} \label{eq:orb_hess}
	\begin{split} 
		\pdv[2]{E}{\kappa_{pq}}{\kappa_{rs}} &=
		4 (P_{pr} - P_{ps} + P_{qs} - P_{qr}) (V_{qrps} + V_{qspr})  
		\\
		&+ 4 (D_{pr} - D_{ps} + D_{qs} - D_{qr}) (4 V_{qprs} - V_{qrps} - V_{qsrp})  
		\\
		& + \delta_{pr} \qty[(4 n_p - 2 n_q - 2 n_s) h_{qs} + W_{pqs}] 
		\\
		& + \delta_{qs} \qty[(4 n_q - 2 n_p - 2 n_r) h_{pr} + W_{qpr}]  
		\\
		& - \delta_{ps} \qty[(4 n_p - 2 n_q - 2 n_r) h_{qr} + W_{pqr}]
		\\
		& - \delta_{qr} \qty[(4 n_q - 2 n_p - 2 n_s) h_{ps} + W_{qps}]
	\end{split}
	\end{equation}
	are both computed from the RDM elements. The Hessian is efficiently computed from the intermediates
	\begin{equation}
	\begin{split}
		W_{pqr} 
		& = 2 \sum_s (2 P_{ps} - P_{qs} - P_{rs}) V_{qssr} 
		\\
		& + 2 \sum_s (2 D_{ps} - D_{qs} - D_{rs}) (2 V_{qrss} - V_{qssr})
	\end{split}
	\end{equation}
	which requires $\order{N^3}$ storage. The mixed elements of the Hessian occur in three types. For the first, the VBS gap $\omega_{\gamma}$ does not correspond to one of the orbitals being rotated
	\begin{equation}
		\qty(\omega^2_{\gamma} + 1 )^{\frac{3}{2}} \pdv[2]{E}{\omega_{\gamma}}{\kappa_{\alpha_\mu \beta_\nu}} =
		2 \sum_\lambda (-1)^\lambda \qty(2 V_{\gamma_\lambda \gamma_\lambda \alpha_\mu \beta_\nu} - V_{\gamma_\lambda \beta_\nu \alpha_\mu \gamma_\lambda} )
		\qty( n_{\alpha_\mu} - n_{\beta_\nu} )
	\end{equation}
	while the remaining two types are
	\begin{subequations}
		\begin{align}
			\begin{split}
				\qty(\omega^2_{\alpha} + 1 )^{\frac{3}{2}} \pdv[2]{E}{\omega_{\alpha}}{\kappa_{\alpha_\mu \beta_\nu}} &=
				2 \omega_{\alpha} V_{ \alpha_{1-\mu} \alpha_\mu \alpha_{1-\mu} \beta_\nu }  		
				+ 2 (-1)^\mu 
				\qty[Q_{\alpha_\mu \beta_\nu} + V_{\alpha_{\mu} \alpha_{\mu} \alpha_{\mu} \beta_{\nu}}]
				\\
				&- 2 \sum_\lambda \qty( 2 V_{\alpha_\lambda \alpha_\lambda \alpha_\mu \beta_\nu} - V_{\alpha_\lambda \beta_\nu \alpha_\mu \alpha_\lambda} )
				\qty[ (-1)^\mu n_{\alpha_{\lambda}} + (-1)^\lambda n_{\beta_\nu}  ]
			\end{split}
			\\
			\begin{split}
				\qty(\omega^2_{\beta} + 1 )^{\frac{3}{2}} \pdv[2]{E}{\omega_{\beta}}{\kappa_{\alpha_\mu \beta_\nu}} &= 
				- 2 \omega_{\beta} V_{ \beta_{1-\nu} \alpha_\mu \beta_{1-\nu} \beta_\nu  }  
				- 2 (-1)^\nu \qty[ Q_{\alpha_\mu \beta_\nu} + V_{\beta_{\nu} \beta_{\nu} \beta_{\nu} \alpha_{\mu} } ]
				\\
				&+ 2 \sum_\lambda \qty( 2 V_{\beta_\lambda \beta_\lambda \alpha_\mu \beta_\nu} - V_{\beta_\lambda \beta_\nu \alpha_\mu \beta_\lambda} )
				\qty[
				(-1)^\lambda  n_{\alpha_{\mu}} + (-1)^\nu n_{\beta_{\lambda}}
				].
			\end{split}
		\end{align}
	\end{subequations}
	The full Hessian can be constructed with $\order{N^4}$ operations so that the scaling bottleneck for orbital-optimized PP (OO-PP) will be, in principle, the $\order{N^5}$ transformation of the two-electron integrals. \cite{Frisch_1990} The orbital gradient \eqref{eq:orb_grad} and the orbital Hessian \eqref{eq:orb_hess} are valid for any seniority-zero wavefunction provided the density matrices $n$, $D$, and $P$. While in this contribution we consider PP only in the valence, core ($n_i=1$) and virtual ($n_a=0$) orbitals may be added without too much difficulty. Density-density 2-RDM elements involving core and virtual orbitals factor into products of occupation numbers [see Eq.~\eqref{eq:GVB_2dm_D}] while pair-transfer elements vanish. The orbital gradient \eqref{eq:orb_grad} will therefore vanish for core-core and virtual-virtual rotations, making PP invariant to such transformations.
	
	Rather than building the complete Hessian, stationary equations are available for the orbital coefficients for the antisymmetrized product of strongly-orthogonal geminals (APSG), a more general structure than PP.\cite{kutzelnigg:1964,hapka:2022} This is the traditional approach for PP \cite{wang:2019,zou:2020} which also employs two coefficients, and thus a constraint, for each pair rather than the VBS gaps $\{\omega_{\alpha}\}$. In the end, the scaling is the same, the VBS gaps remove redundancies and have clear physical meanings, and the complete Hessian is computable. We feel the current approach is more transparent.

	Variational OO-PP is a \emph{mean-field of non-interacting pairs} and the model Hamiltonian \eqref{eq:H_GVB} provides a basis for the Hilbert space to account for the outstanding weak correlation (relative to PP). In the seniority-zero sector, $\ket{\PP}$ couples to three types of excitations through $\hH_\text{C}$, as shown in Fig.~\ref{fig:sketch}. In each VBS $\alpha$, the process $\ket{\alpha_{-}}\rightarrow \ket{\alpha_{+}}$ is called an \emph{intra-pair excitation} and denoted $\ket*{\PP^\alpha_\alpha}$. An \emph{inter-pair excitation} $\ket*{\PP^\beta_\alpha}$ from $\alpha$ to $\beta$ refers to $\ket{\alpha_{-}\beta_{-}}\rightarrow \ket{\ger{\beta}\ung{\beta}}$. Finally, a \emph{double pair excitation}$\ket*{\PP^{\alpha\beta}_{\alpha\beta}}$ refers to $\ket{\alpha_{-}\beta_{-}}\rightarrow \ket{\alpha_{+}\beta_{+}}$. Additional couplings were possible for RG states, while they are totally forbidden for PP. A CI treatment of PP with its singles and doubles (CISD) is possible, though in Ref.~\onlinecite{johnson:2024b} it was found for RG states that EN2 was much more attractive: building the RG-based CISD matrix cost $\order{N^8}$ operations while EN2 cost only $\order{N^6}$. For the dissociation of linear H$_8$, the maximum disagreement between the two methods was about \SI{2d-5}{\hartree} for very short interatomic distances, and this disagreement went to zero quickly as the distance increased. In any case, the CISD treatment of PP was performed to verify our expressions were correct.
	
	Here, an EN reference Hamiltonian can be written
	\begin{equation}
	\begin{split}
		\hH_0 &= \dyad{\PP}{\PP} \hH_\text{C} \dyad{\PP}{\PP}
		\\
		& + \sum_{\alpha\beta} \dyad*{\PP^\beta_\alpha}{\PP^\beta_\alpha} \hH_\text{C} \dyad*{\PP^\beta_\alpha}{\PP^\beta_\alpha}
		\\
		& + \sum_{\alpha<\beta} \dyad*{\PP^{\alpha\beta}_{\alpha\beta}}{\PP^{\alpha\beta}_{\alpha\beta}} \hH_\text{C} \dyad*{\PP^{\alpha\beta}_{\alpha\beta}}{\PP^{\alpha\beta}_{\alpha\beta}}
	\end{split}
	\end{equation}
	with internal space $\ket{\PP}$ and external space built from intra-pair, inter-pair, and double-pair excitations (see Fig.~\ref{fig:sketch}). The first-order correction to the energy is zero, while the second-order correction is
	\begin{equation} \label{eq:enpt2_energy}
		E^{(2)} 
		= \sum_{\alpha\beta} \frac{ \abs*{ \mel*{\PP^\beta_\alpha}{\hH_\text{C}}{\PP} }^2 }{E^{(0)} - E_{\alpha}^{\beta}}
		+ \sum_{\alpha<\beta} \frac{ \abs*{ \mel*{\PP^{\alpha\beta}_{\alpha\beta}}{\hH_\text{C}}{\PP} }^2 }{E^{(0)} - E_{\alpha\beta}^{\alpha\beta}}
	\end{equation}
	where $E^{(0)} = \mel*{\PP}{\hH_\text{C}}{\PP}$, $E_{\alpha}^{\beta} = \mel*{\PP^\beta_\alpha}{\hH_\text{C}}{\PP^\beta_\alpha}$, and $E_{\alpha\beta}^{\alpha\beta} = \mel*{\PP^{\alpha\beta}_{\alpha\beta}}{\hH_\text{C}}{\PP^{\alpha\beta}_{\alpha\beta}}$.
	This correction is simplified in Appendix \ref{sec:pt2}, leading to an evaluation of Eq.~\eqref{eq:enpt2_energy} with $\order{N^2}$ operations, essentially free once the OO-PP mean-field is constructed. Intra-pair excitations do not contribute at all: as shown in Appendix \ref{sec:pt2}, the transition $\mel{\PP^{\alpha}_{\alpha}}{\hH_\text{C}}{\PP}$ is equivalent to the gradient condition \eqref{eq:grad_vanish} and therefore vanishes. This is analogous to the Brillouin theorem within HF theory. \cite{szabo_book} This EN2 treatment only includes the seniority-zero sector as the goal is to compare with pCCD. An EN2 correction including PP excited states of non-zero seniorities will be computed for a future contribution in the same manner as for RG states. \cite{johnson:2025b,johnson:2025c}
	
	The natural orbital functions of Piris are also directly related to PP. At half-filling, the PNOF5\cite{piris:2011} functional is equivalent to PP. It is otherwise equivalent to APSG.\cite{pernal:2013} For strong correlation, the best-performing functional is PNOF7\cite{piris:2017,mitxelena:2018} which can be seen as an update to PNOF5:
	\begin{align}
		E_\text{PNOF7} = E_\text{PNOF5} + \Delta E_\text{PNOF7}
	\end{align}
	Usually, PNOF7 is treated as a mean-field, solved with Lagrange multipliers in an SCF-like procedure, but PNOF7 is not $N$-representable. PNOF5, being equivalent to PP, is $N$-representable and variational. Thus, PNOF7 could be computed as a perturbative update to the variational PNOF5. In the present language, this update becomes
	\begin{align}
		\Delta E_\text{PNOF7} &= - \frac{1}{4}
		\sum_{\alpha} \sum_{\beta (\neq \alpha)} 
		\frac{1}{(\omega_{\alpha}^2 + 1)^{\frac{1}{2}}(\omega_{\beta}^2 + 1)^{\frac{1}{2}}}
		\sum_{\mu} \sum_{\nu} L_{\alpha_{\mu} \beta_{\nu}}.
	\end{align}
	
	\subsection{Pair Coupled-Cluster Doubles}
	We now move on to describe the connection with pCCD.
	The pCCD energy is very similar to the EN2-corrected PP energy. 
	This can be evidenced by examining the pCCD wavefunction ansatz, which reads, in its most conventional form, 
	\begin{equation}
		\ket{\pCCD} 
		= e^{\hT} \ket{\SD} 
	\end{equation}
	where
	\begin{equation}
	 	\hT = \sum_{ia} t^a_i \Sp{a} \Sm{i}
	\end{equation}
	is the cluster operator (which is restricted to paired double excitations at the pCCD level) and
	\begin{equation}
		\ket{\SD} = \prod_{\alpha} \ket{\ger{\alpha}}
	\end{equation}
	is the closed-shell ground-state Slater determinant in which all the bonding orbitals are doubly occupied. 
	Again, as the orbitals are optimized from the bifunctional expression of the pCCD energy, the Slater determinant of bonding orbitals is \emph{not} the HF Slater determinant.
	The pCCD state can be recast
	\begin{equation}
	\begin{split}
		\ket{\pCCD} 
		& = \exp( \sum_{\alpha\beta} t_{\alpha}^{\beta} \Sp{\ung{\beta}} \Sm{\ger{\alpha}} ) \ket{\SD} 
		\\
		& = \exp( \sum_{\beta(\neq \alpha)} t_{\alpha}^{\beta} \Sp{\ung{\beta}} \Sm{\ger{\alpha}} ) \exp( \sum_{\alpha} t_{\alpha}^{\alpha} \Sp{\ung{\alpha}} \Sm{\ger{\alpha}} ) \ket{\SD}
	\end{split}
	\end{equation}
	in terms of amplitudes $t_{\alpha}^{\beta}$ to be determined from a set of non-linear equations obtained by projection of the Schr\"odinger equation over the excited Slater determinants $\ket*{\SD_\alpha^\beta} = \Sp{\ung{\beta}} \Sm{\ger{\alpha}} \ket{\SD}$ from the left, as follows:
	\begin{equation} \label{eq:amp_eq_right}
		\mel{\SD_\alpha^\beta}{\stH}{\SD} = 0
	\end{equation}
	where $\stH = e^{-\hT} \hH_\text{C} e^{\hT}$ is the similarity-transformed Hamiltonian.
	Splitting the exponential into diagonal and off-diagonal pieces is possible and makes the individual contributions evident: the diagonal cluster amplitudes will turn $\ket{\SD}$ into $\ket{\PP}$, that is,
	\begin{equation} \label{eq:pp_pccd_diag}
	\begin{split}
		\ket{\PP} 
		& = \exp( \sum_{\alpha} t_\alpha^\alpha \Sp{\ung{\alpha}} \Sm{\ger{\alpha}} ) \ket{\SD}
		\\
		& = \prod_{\alpha} \exp( t_\alpha^\alpha \Sp{\ung{\alpha}} \Sm{\ger{\alpha}} ) \ket{\SD}
		\\
		& = \prod_{\alpha} \qty(1 + t^{\alpha}_{\alpha} \Sp{\ung{\alpha}}\Sm{\ger{\alpha}} ) \ket{\SD}
	\end{split}
	\end{equation}
	while the off-diagonal cluster amplitudes will add the inter-pair excitations. Double pair excitations will be approximated as products of the cluster amplitudes. 

	First, the diagonal cluster amplitudes transform $\ket{\SD}$ into $\ket{\PP}$. From Eq.~\eqref{eq:amp_eq_right}, the (quadratic) amplitude equations for the right amplitudes $\{ t_i^a \}$ are \cite{henderson:2014b}
	\begin{equation}
	\label{eq:t_amp}
 	\begin{split}
		0 & = L_{ai} + 2\qty( f_{aa} - f_{ii} - \sum_j L_{ja} t_j^a - \sum_b L_{ib} t_i^b ) t_i^a 
		\\
		& - 2 \qty( 2 J_{ia} - K_{ia} - L_{ia} t_i^a ) t_i^a 
		+ \sum _b L_{ba} t_i^b + \sum_j L_{ji} t_j^a + \sum_{jb} L_{jb} t_j^a t_i^b
	\end{split}
 	\end{equation}
	where $f_{pq}$ are the elements of the Fock operator.
	In the single-pair approximation, it reduces to the following simple form:
	\begin{equation} \label{eq:quadratic_t}
		1 + 2 \omega_{\alpha} t_{\alpha}^{\alpha} - (t_{\alpha}^{\alpha})^2 = 0
 	\end{equation}
	with
	\begin{equation}
	\begin{split}
		\omega_{\alpha} 
		& = \frac{f_{\ung{\alpha}\ung{\alpha}} - f_{\ger{\alpha}\ger{\alpha}} - 2 J_{\ger{\alpha}\ung{\alpha}} + K_{\ger{\alpha}\ung{\alpha}} + \frac{L_{\ger{\alpha}\ger{\alpha}}+L_{\ung{\alpha}\ung{\alpha}}}{2}}{L_{\ger{\alpha}\ung{\alpha}}}
		\\
		& = \frac{h_{\ung{\alpha}\ung{\alpha}} + \frac{L_{\ung{\alpha}\ung{\alpha}}}{2} - h_{\ger{\alpha}\ger{\alpha}} - \frac{L_{\ger{\alpha}\ger{\alpha}}}{2}}{L_{\ger{\alpha}\ung{\alpha}}}
	\end{split}
	\end{equation}
	which matches the expression obtained from Eq.~\eqref{eq:grad_vanish} in the single-pair approximation.
	The ground-state amplitude for the pair $\alpha$ is thus simply given by
	\begin{equation} \label{eq:t_pCCD}
		t_{\alpha}^{\alpha} 
		= \omega_{\alpha} - \sqrt{1 + \omega_{\alpha}^2}
		= - \sqrt{ \frac{n_{\ung{\alpha}}}{n_{\ger{\alpha}}} }, 
	\end{equation}
	which corresponds to the state $\ket{\alpha_-}$. 
	As Eq.~\eqref{eq:quadratic_t} is quadratic, there is evidently a second root corresponding to the state $\ket{\alpha_+}$.
	The pCCD energy is defined via the following energy bi-functional \cite{henderson:2014b}
	\begin{equation}
		E_\text{pCCD} = \mel{\SD}{(1 + \hZ) \stH}{\SD}
	\end{equation}
	where $\hZ = \sum_{ia} z_a^i \Sp{i} \Sm{a}$ is a de-excitation operator.
	The (linear) amplitude equations for the right amplitudes $\{ z^i_a \}$ are \cite{henderson:2014b}
	\begin{equation}
	\label{eq:z_amp}
	\begin{split}
		0 & = L_{ia} + 2\qty(f_{aa} - f_{ii} - \sum_j L_{ja} t_j^a - \sum_b L_{ib} t_i^b ) z^i_a 
		 - 2\qty( 2 J_{ia} - K_{ia}- 2 L_{ia} t_i^a ) z^i_a 
		\\
		& + \sum _b L_{ab}  z^i_b + \sum_j L_{ij} z^j_a + \sum_{jb} t_j^b \qty(L_{ib} z^j_a + L_{ja} z^i_b )
		- 2 L_{ia} \qty( \sum_j z^j_a t_j^a + \sum _b z^i_b t_i^b )
	\end{split}
	\end{equation}
	In the single-pair approximation, it reduces to the following simple form:
	\begin{equation}
		1 + 2 \qty(\omega_{\alpha} - t_{\alpha}^{\alpha}) z_{\alpha}^{\alpha} = 0.
 	\end{equation}	
	The left ground-state amplitude for the pair $\alpha$ is thus simply given by
	\begin{equation}
		z_{\alpha}^{\alpha} 
		= - \frac{1}{2\qty(\omega_{\alpha} - t_{\alpha}^{\alpha})}
		= - \sqrt{n_{\ger{\alpha}}n_{\ung{\alpha}}}
 	\end{equation}	
	where $t_{\alpha}^{\alpha}$ is given by Eq.~\eqref{eq:t_pCCD}.
	The left and right amplitudes are then employed to compute the 1- and 2-RDM and, eventually, the energy.
	For example, the non-zero elements of the 1-RDM (which is diagonal at the pCCD level) are
	\begin{subequations}
	\begin{align}
		n_{i} & = \qty( 1 - \sum_a t_i^a z_a^i )
		\\ 
		n_{a} & = \sum_i t_i^a z_a^i.
	\end{align}
	\end{subequations}
	In the single-pair approximation, it is easily shown that one recovers the expressions of Eq.~\eqref{eq:GVB_1dm}.
	Likewise, for the elements of the 2-RDM $P$ (hence the energy), while the elements of $D$ vanish in the single-pair approximation.
	This slightly generalizes Cullen's derivation \cite{cullen:1996} and shows that, within the single-pair approximation, pCCD and PP are strictly equivalent.
	However, it does not demonstrate the equivalence of these two formalisms beyond the single-pair approximation.

	\section{Numerical Results} \label{sec:num}
	To illustrate the theoretical results discussed above, we consider chains of (equidistant) Hydrogen atoms of increasing size at half-filling, which here means the minimal basis STO-6G. Of course, PP can include cores and virtuals, but that is not the present purpose. As OO-PP, OO-PP-EN2, and OO-pCCD are seniority-zero states, results will be compared with OO-DOCI. In all cases, we report \emph{reduced energies}, i.e., the energy divided by the number of electrons.

	OO-PP results were computed using the Newton-Raphson algorithm by constructing the full Hessian. Symmetric and antisymmetric linear combinations of the 1s hydrogen orbitals on neighbouring sites served as a sufficient initial guess for the orbitals, while the VBS gaps were initialized to a value smaller than one. Generally, for distances larger than $R_{\ce{H-H}}=\SI{2.5}{\bohr}$, this approach converges in 3-4 iterations. At shorter distances, the Hessian develops negative eigenvalues and we adopt a level-shift approach so that the Hessian remains positive definite. The time spent computing the EN2 correction is negligible in comparison. The pCCD calculations have been performed using the same methodology as in Ref.~\onlinecite{rodriguez-mayorga:2025} with the \texttt{MOLGW} program \cite{bruneval2016molgw} that incorporates the stand-alone NOFT module \cite{rodriguez2022snoft} based on the \texttt{DoNOF} program. \cite{piris2021donof} 
		
	RHF and FCI results were obtained with PySCF, \cite{sun:2018} while OO-DOCI results were computed with our own code. OO-DOCI convergence is difficult near the minimum: the optimization is decoupled, switching between exact diagonalization of the DOCI Hamiltonian matrix and Newton-Raphson optimization of the orbital coefficients. More robust and efficient optimization strategies have been proposed in the literature.\cite{siegbahn:1981,werner:1985,yao:2021}
	
\begin{figure*}
		\includegraphics[width=\linewidth]{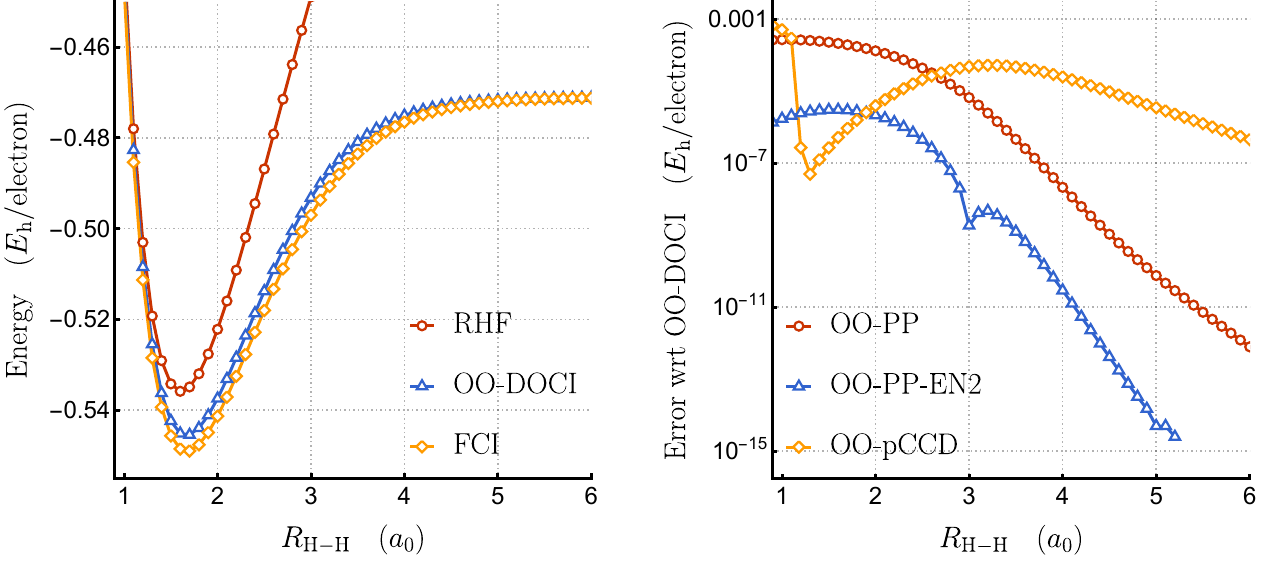}
		\caption{Symmetric dissociation of \ce{H4} chain in STO-6G. Orbitals were optimized separately for PP, pCCD, and DOCI. 
		Left: Reduced energies for RHF, OO-DOCI, and FCI. 
		Right: Absolute errors of OO-PP, OO-PP-EN2, and OO-pCCD with respect to OO-DOCI plotted logarithmically. 
		OO-PP-EN2 is above OO-DOCI until $R_{\ce{H-H}}=\SI{3.1}{\bohr}$, where it goes below and overcorrelates the rest of the way to dissociation. 
		OO-pCCD always overcorrelates relative to OO-DOCI.}
		\label{fig:h4_energy}
	\end{figure*}

	First, we consider the smallest non-trivial chain \ce{H4} and the results are presented in Fig.~\ref{fig:h4_energy}. 
	It is difficult to visually discern the reduced energies of OO-PP, OO-PP-EN2, OO-pCCD, and OO-DOCI, so only the OO-DOCI curve is plotted in the left panel of Fig.~\ref{fig:h4_energy}. 
	Indeed, one can see in the right panel of Fig.~\ref{fig:h4_energy} that the respective errors, in terms of reduced energies, for the three approximate methods are below \SI{1}{\milli\hartree} per electron everywhere. 
	As OO-PP is a variational approximation, it is always above OO-DOCI. 
	OO-PP-EN2 is above OO-DOCI until $R_{\ce{H-H}}=\SI{3.1}{\bohr}$, where it falls below OO-DOCI. 
	OO-pCCD is always below OO-DOCI. 
	At short distances, OO-pCCD performs the best, but is quickly overtaken by OO-PP-EN2 and even OO-PP. 
	As it has an exponential parametrization, pCCD is extensive, as is PP due to Eq.~\eqref{eq:pp_pccd_diag}. The OO-PP-EN2 correction goes to zero rapidly for distinct subsystems as the optimal orbitals are localized.\cite{limacher:2014a} A detailed analysis of the extensivity of the EN2 correction is presented in Ref.~\onlinecite{johnson:2024b}.
	Evidently, the energy differences among these methods are very small and are unlikely to be significant in practical applications.	
	
	To better demonstrate how small the energy differences between the methods are, reduced energies for PP, EN2, pCCD, and DOCI were computed in the OO-PP, OO-pCCD, and OO-DOCI orbitals. 
	The reduced energies at $R_{\ce{H-H}}=\SI{1.6}{\bohr}$, near equilibrium, are shown in Table \ref{tab:short}. 
	This point is chosen as it represents where weak correlation is dominant. 
	Here, EN2 and pCCD are identical to DOCI. 
	PP is a bit above, as it misses the weak correlation.
	
	\begin{table}[h!]
		\caption{Reduced energies computed at $R_{\ce{H-H}}=\SI{1.6}{\bohr}$, near the equilibrium point. Each column represents a set of orbitals while each row is a method. All results computed in STO-6G.}
		\label{tab:short}
		\centering
		\begin{tabular}{C{2cm} C{2cm} C{2cm} C{2cm}}
			\toprule
			& \multicolumn{3}{c}{\textbf{Orbital set}} \\
			\cline{2-4}
			\textbf{Method} & \textbf{PP} & \textbf{pCCD} & \textbf{DOCI} \\
			\toprule
			\textbf{PP}    & -0.54492 & -0.54492 & -0.54492 \\
			\textbf{PP-EN2} & -0.54513 & -0.54513 & -0.54513 \\
			\textbf{pCCD}  & -0.54513 & -0.54513 & -0.54513 \\
			\textbf{DOCI}  & -0.54513 & -0.54513 & -0.54513 \\
			\bottomrule
		\end{tabular}
	\end{table}
	
	Reduced energies for each method in each set of orbitals at $R_{\ce{H-H}}=\SI{3.2}{\bohr}$ are reported in Table \ref{tab:long}. 
	At this point, the VBS gaps are both less than 1, indicating that the Coulomb repulsion is larger than the bonding/antibonding energy splitting.
	Here, strong correlation is dominant, and pCCD overcorrelates on the order of \SI{0.05}{\milli\hartree\per electron} while the other methods are indiscernible.
	
	\begin{table}[h!]
		\caption{Reduced energies computed at $R_{\ce{H-H}}=\SI{3.2}{\bohr}$. Each column represents a set of orbitals, while each row is a method. All results computed in STO-6G.}
		\label{tab:long}
		\centering
		\begin{tabular}{C{2cm} C{2cm} C{2cm} C{2cm}}
			\toprule
			& \multicolumn{3}{c}{\textbf{Orbital set}} \\
			\cline{2-4}
			\textbf{Method} & \textbf{PP} & \textbf{pCCD} & \textbf{DOCI} \\
			\toprule
			\textbf{PP}    & -0.48728 & -0.48728 & -0.48728 \\
			\textbf{PP-EN2} & -0.48728 & -0.48728 & -0.48728 \\
			\textbf{pCCD}  & -0.48733 & -0.48733 & -0.48733 \\
			\textbf{DOCI}  & -0.48728 & -0.48728 & -0.48728 \\
			\bottomrule
		\end{tabular}
	\end{table}
	
	Reduced energies for each method in each set of orbitals at $R_{\ce{H-H}}=\SI{2.4}{\bohr}$ are reported in Table \ref{tab:mid}. 
	This point lies halfway between the minimum and the point where strong correlation starts to dominate, and thus should represent a case where both weak and strong correlation are important. 
	As expected, PP misses weak correlation and thus is too high. 
	Both EN2 and pCCD are essentially the same as DOCI in all cases.
	
	\begin{table}[h!]
		\caption{Reduced energies computed at $R_{\ce{H-H}}=\SI{2.4}{\bohr}$. Each column represents a set of orbitals, while each row is a method. All results computed in STO-6G.}
		\label{tab:mid}
		\centering
		\begin{tabular}{C{2cm} C{2cm} C{2cm} C{2cm}}
			\toprule
			& \multicolumn{3}{c}{\textbf{Orbital set}} \\
			\cline{2-4}
			\textbf{Method} & \textbf{PP} & \textbf{pCCD} & \textbf{DOCI} \\
			\toprule
			\textbf{PP}    & -0.51856 & -0.51856 & -0.51856 \\
			\textbf{PP-EN2} & -0.51862 & -0.51862 & -0.51862 \\
			\textbf{pCCD}  & -0.51863 & -0.51864 & -0.51864 \\
			\textbf{DOCI}  & -0.51862 & -0.51862 & -0.51862 \\
			\bottomrule
		\end{tabular}
	\end{table}
	
\begin{figure*}
		\includegraphics[width=\linewidth]{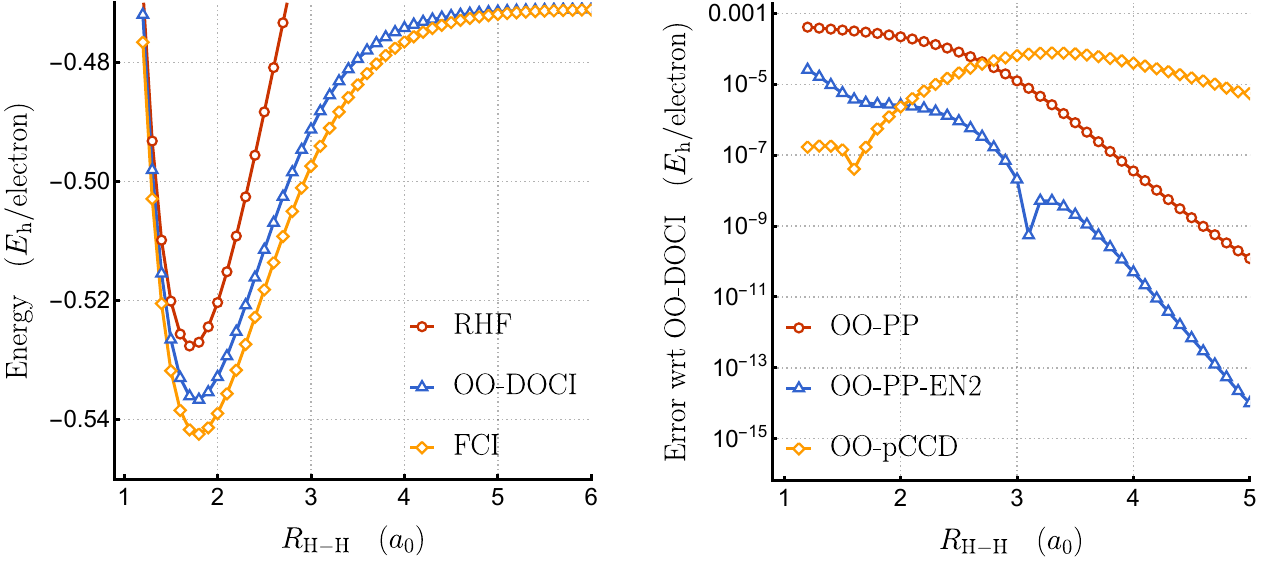}
		\caption{Symmetric dissociation of \ce{H10} chain in STO-6G. 
			Orbitals were optimized separately for PP, pCCD, and DOCI. 
			Left: Reduced energies for RHF, OO-DOCI, and FCI. 
			Right: Absolute errors of OO-PP, OO-PP-EN2, and OO-pCCD with respect to OO-DOCI plotted logarithmically. 
			OO-PP-EN2 is above OO-DOCI until $R_{\ce{H-H}}=\SI{3.2}{\bohr}$ where it goes below and overcorrelates the rest of the way to dissociation. 
			OO-pCCD always overcorrelates relative to OO-DOCI.}
		\label{fig:h10_energy}
	\end{figure*}

	Reduced energies for \ce{H10} are shown in Fig.~\ref{fig:h10_energy}, and the results are qualitatively the same as for \ce{H4}. 
	Notice that OO-DOCI is further away from FCI than it was for \ce{H4} as there are more weak correlation effects from the seniority-two and seniority-four sectors. 
	It is again difficult to discern OO-PP, OO-PP-EN2, and OO-pCCD from OO-DOCI, so we plot only the differences in the right panel of Fig.~\ref{fig:h10_energy}.
		
\begin{figure*}
		\includegraphics[width=0.45\linewidth]{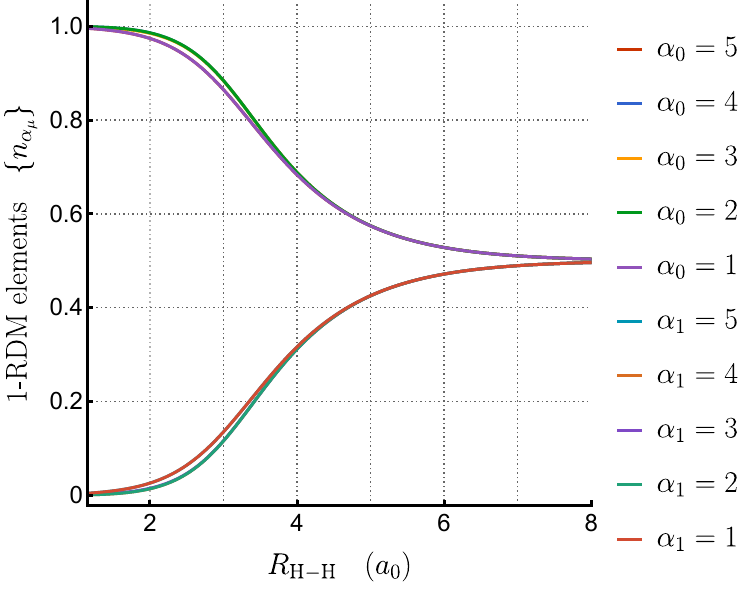}\hfill
		\includegraphics[width=0.45\linewidth]{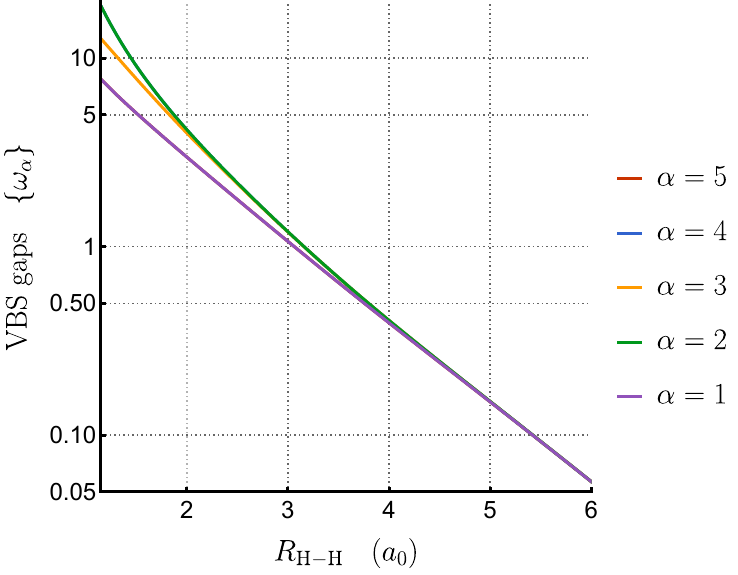}
		\caption{
				Symmetric dissociation of \ce{H10} chain in STO-6G.
				Left: 1-RDM elements $\{n_{\alpha}\}$ for \ce{H10} optimized with OO-PP/STO-6G. 
				Right: VBS gaps $\{\omega_{\alpha}\}$ for \ce{H10} optimized with OO-PP/STO-6G plotted logarithmically. 
				}
		\label{fig:h10_noccs}
	\end{figure*}

	From the OO-PP solution for \ce{H10}, the 1-RDM elements and the corresponding VBS gaps are shown in Fig.~\ref{fig:h10_noccs}. 
	One sees immediately that the dissociations are smooth, though the bonds are not equivalent but break into a 2-2-1 pattern: the two bonds furthest to the edges of the molecule are equivalent, the two bonds that are one bond in from the edges are equivalent, and the bond at the centre of the molecule is non-degenerate. The VBS gaps decay exponentially with $R_{\ce{H-H}}$, consistent with previous observations for RG states.\cite{johnson:2024b} In other words, it is reasonable to consider
	\begin{align} \label{eq:vbs_r}
		\omega_{\alpha} = B_{\alpha} \exp \left( - A_{\alpha} R \right),
	\end{align}
	with only slight deviations at small $R_{\ce{H-H}}$. Near equilibrium, the VBS gaps $\{\omega_{\alpha}\}$ are large and the pairs of electrons lie principally in the bonding orbital (weak electron correlation). At dissociation, the VBS gaps $\{\omega_{\alpha}\}$ go to zero and the occupations of the bonding and antibonding orbitals are the same (strong electron correlation). These two effects are in balance when $\omega_{\alpha}=1$, and we have argued previously for RG states that this point marks the transition from weak to strong electron correlation for the particular pair of electrons.\cite{johnson:2024b} For PP, the pair occupation numbers are explicit functions of the VBS gaps, giving the specific value for the bonding orbital occupation as $\frac{1}{2}\left(1 + \frac{1}{\sqrt{2}}\right) \approx 0.8536$.
	
\begin{figure*}
		\includegraphics[width=\linewidth]{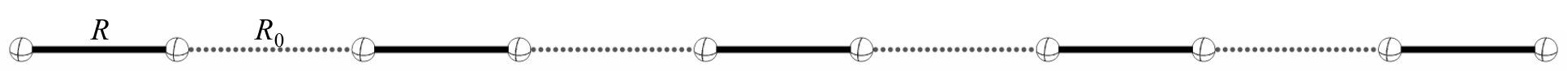}
		\caption{
			Progressive dimerization due to Peierls instability of \ce{H10}: $R_0$ is kept fixed at 3.6 $a_0$ while the distance $R$ is varied.
			}
		\label{fig:d10_struct}
	\end{figure*}

	While equidistant chains of hydrogen atoms are standard as a test for strong electron correlation, they are inherently unphysical: at short distances, these systems do not remain equidistant, but spontaneously dimerize into \ce{H2} molecules, a mechanism known as the Peierls instability.\cite{peierls_book} To describe this instability, we consider a system of hydrogen atoms as in Fig.~\ref{fig:d10_struct}.
	Here, the distances between hydrogen atoms 2 and 3 (and so on) are kept fixed at $R_0=3.6\;a_0$ while the distance $R$ between hydrogens 1 and 2 is varied.
	\begin{figure*}
		\includegraphics[width=0.45\linewidth]{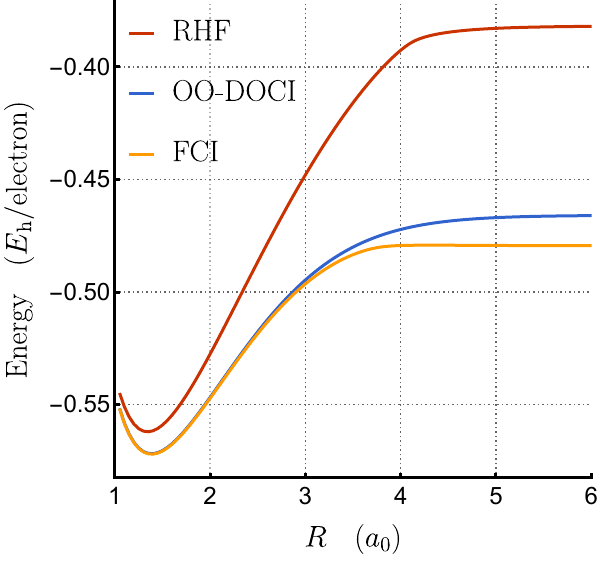}\hfill
		\includegraphics[width=0.45\linewidth]{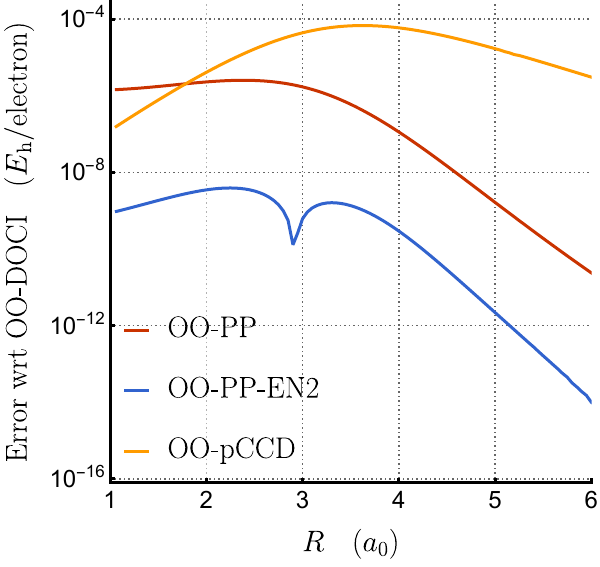}
		\caption{Progressive dimerization of \ce{H10} in STO-6G. 
			Left: Reduced energies for RHF, OO-DOCI, and FCI. 
			Right: Absolute errors of OO-PP, OO-PP-EN2, and OO-pCCD with respect to OO-DOCI plotted logarithmically. 
			OO-PP-EN2 is above OO-DOCI until $R=\SI{2.9}{\bohr}$ where it goes below and overcorrelates the rest of the way to dissociation. 
			OO-pCCD always overcorrelates relative to OO-DOCI.}
		\label{fig:d10_energy}
	\end{figure*}
	From the numerical results in Fig.~\ref{fig:d10_energy} it is evident that the dimerization does indeed stabilize the energy. OO-DOCI is now a much better treatment at the equilibrium distance, while OO-PP, OO-PP-EN2, and OO-pCCD all deviate smoothly from OO-DOCI. At long distances, OO-DOCI is further away from FCI as the system is trying to dimerize in the opposite manner with individual hydrogen atoms at each end. The curves are parallel, however, and including seniorities two and four would correct the problem.
	
\begin{figure*}
		\includegraphics[width=0.45\linewidth]{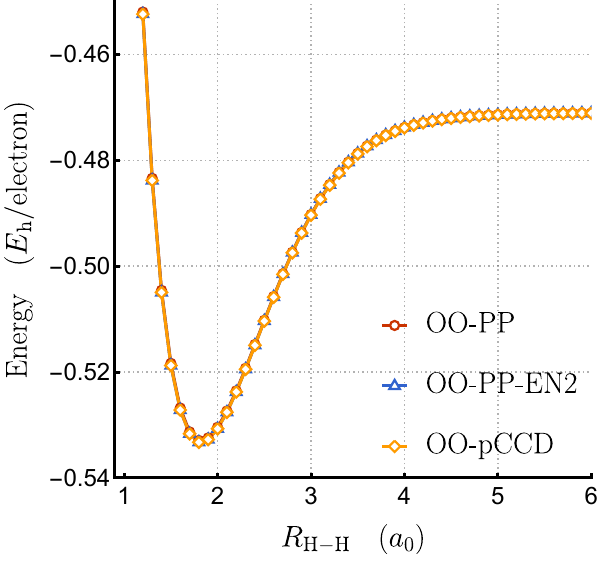}
		\caption{Symmetric dissociation of \ce{H50} chain in STO-6G. 
		Orbitals were optimized separately for PP and pCCD. 
		OO-PP-EN2 is indiscernible from OO-pCCD.}
		\label{fig:h50_energy}
	\end{figure*}

	Finally, OO-PP, OO-PP-EN2, and OO-pCCCD were computed for equidistant \ce{H50} and the results are shown in Fig.~\ref{fig:h50_energy}. 
	As \ce{H50} is too large for DOCI and FCI, the take-home message is that both OO-PP-EN2 and OO-pCCD remain quite feasible and give essentially the same result.
	Rather than plotting the individual 1-RDM elements for \ce{H50}, we will take the opportunity to discuss reduced quantities summarizing the physical behaviour, which in condensed-matter theory are known as \emph{order parameters}. One advantage of seeing PP as an eigenvector of Eq.~\eqref{eq:H_GVB}, rather than a wavefunction ansatz, is that derived quantities are simple functions of the VBS gaps. First, define the bond order for each VBS as
	\begin{align}
		\chi_{\alpha} = n_{\alpha_0} - n_{\alpha_1} = \frac{\omega_{\alpha}}{\sqrt{\omega^2_{\alpha} + 1}}
	\end{align}
	half the number of bonding electrons minus the number of antibonding electrons. Here, $0 \leq \chi_{\alpha} < 1$, with a value approaching 1 in the limit of large $\omega_{\alpha}$. In spin systems, this is the \emph{magnetization}, from which thermodynamic quantities are computed.\cite{baxter_book}	
	Another order parameter, the Jaynes entropy\cite{jaynes:1957a,jaynes:1957b}
	\begin{align} \label{eq:jaynes}
		S_J = - \sum_p n_p \ln n_p
	\end{align}
	is understood to be indicative of strong correlation in bond-breaking processes.\cite{collins:1993,wang:2022,zamani:2025} However, the Jaynes entropy gives \emph{no} description of weak correlation,\cite{jerzy:2024} as occupation numbers close to 0 or 1 give vanishing contributions in Eq.~\eqref{eq:jaynes}. For PP, the Jaynes entropy simplifies to a sum of individual contributions from each VBS
	\begin{align}
		S_{\alpha} = \ln 2 - \sum^{\infty}_{n=1} \frac{\chi^{2n}_{\alpha}}{2n(2n-1)},
	\end{align}
	in terms of the bond order. Notice that since $ \abs{\chi_{\alpha}} < 1$, this series always converges.
	
\begin{figure*}
		\includegraphics[width=0.45\linewidth]{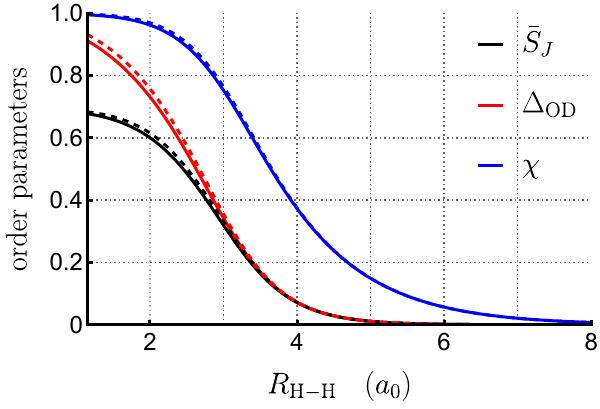}
		\caption{Order parameters in the symmetric dissociation of equidistant \ce{H10} (solid lines) and \ce{H50} (dashed lines) chains in STO-6G, computed from the OO-PP RDM elements.}
		\label{fig:h50_order}
	\end{figure*}

	Finally, the off-diagonal long-range order (ODLRO),\cite{penrose:1956,yang:1962,bloch:1965} defined as
	\begin{align}
		\Delta_\text{OD} = \frac{2}{N} \sum_{pq} P_{pq} = 1 - \frac{2}{N} \sum_{\alpha} \frac{1}{\sqrt{\omega^2_{\alpha} + 1}}
	\end{align}
	is an order parameter describing systems dominated by pairing interactions. Usually, the ODLRO is computed for attractive systems, where it identifies an eigenvalue of $P$ which scales with the size of the system, and hence \emph{long-range order}.\cite{tian:2001,zhou:2002,faribault:2008,faribault:2010} In bond-breaking processes, the pairing is repulsive and the ODLRO goes sharply to zero as seen in Fig.~\ref{fig:h50_order}. The name is thus less descriptive, but the order parameter itself demonstrates the transition in behaviour.

	To demonstrate that these three order parameters contain the same information, we plot 
	\begin{align}
		\chi      &= \frac{2}{N} \sum_{\alpha} \chi_{\alpha}, \\
		\bar{S}_J &= - \frac{2}{N} S_J + \ln 2,
	\end{align}
	and $\Delta_\text{OD}$ for linear equidistant \ce{H10} and \ce{H50} in Fig.~\ref{fig:h50_order}.	
	One notices immediately that the three order parameters are qualitatively the same, and nearly sigmoidal in shape. $\Delta_\text{OD}$ will eventually level out to 1 as the individual $\omega_{\alpha}$ become large. The three order parameters are essentially invariant to size, with minor variations as the bonds in linear hydrogen chains are not all equivalent. Of the three, the bond order is to be preferred as it is the most direct to compute, and is most familiar to chemists.
	
	We will close with an intuitive interpretation from the bond order. As a function of $\omega_{\alpha}$, the bond order $\chi_{\alpha}$ is strictly concave, but as a function of $R$, there is clearly an inflection point. It is not unreasonable to interpret the inflection point as where the bond breaks.\cite{hait:2023} In Ref.~\onlinecite{hait:2023}, the authors mention that this point is usually incredibly difficult to compute, as it would come from an FCI wavefunction, and instead consider the maximum of the polarizability along the bond axis as the point where the bond breaks. One can however easily calculate this point for PP, using
	\begin{align}
		\frac{\partial^2 \chi_{\alpha}}{\partial R^2} = 
		\frac{\partial^2 \chi_{\alpha}}{\partial \omega_{\alpha}^2} \left( \frac{\partial \omega_{\alpha}}{\partial R} \right)^2
		+ \frac{\partial \chi_{\alpha}}{\partial \omega_{\alpha}} \frac{\partial^2 \omega_{\alpha}}{\partial R^2},
	\end{align}
	with the parametrization \eqref{eq:vbs_r} giving
	\begin{align}
		\frac{1}{A^2_{\alpha}} \left( \omega^2_{\alpha} + 1 \right)^{\frac{5}{2}} \frac{\partial^2 \chi_{\alpha}}{\partial R^2} =
		\omega_{\alpha} \left( 1 - 2 \omega^2_{\alpha} \right),
	\end{align}
	and thus a solution of $\omega_{\alpha} = \frac{1}{\sqrt{2}}$. This corresponds to a bond order of $\chi_{\alpha}=\frac{1}{\sqrt{3}}\approx 0.5774$, or a bonding orbital occupation of $n_{\alpha_0} \approx 0.7887$. These are ideal values: inclusion of weak correlation will decrease them slightly. From Fig.~\ref{fig:h10_noccs}, it is clear that the parametrization \eqref{eq:vbs_r} is not perfect, but it is only needed in the neighbourhood of the inflection point. For \ce{H2} in the minimal basis STO-6G, OO-PP thus predicts that the bond breaks near $R \approx 3.335\;a_0$ which is close to the value $R=1.75$ \AA $\;\approx3.307\;a_0$ obtained in Ref.~\onlinecite{hait:2023} from the polarizability (FCI/aug-cc-pVTZ).		
	
	\section{Conclusion}
	We have demonstrated that the PP wavefunction, long recognized as a key model for strong correlation, naturally emerges as an eigenvector of a simplified reduced BCS Hamiltonian formulated in the space of bonding and antibonding orbitals. This construction clarifies the formal relationship between PP, pCCD, and RG states, establishing a unifying framework that connects orbital- and geminal-based descriptions of electron correlation. Within this picture, the complementary eigenvectors of the simplified Hamiltonian provide a systematic route to include weak correlation effects through perturbative corrections. In particular, EN2 built upon PP yields results nearly indistinguishable from pCCD, confirming that the latter effectively incorporates the same physical content through a CC parameterization. Because PP is derived from a well-defined model Hamiltonian, one can explicitly construct a complete Hilbert space in which corrections may be introduced systematically, much like in single-reference methods for weakly correlated systems. While the present treatment is restricted to the seniority-zero sector, including the PP limit to RG states with non-zero seniority\cite{johnson:2025b,johnson:2025c} is a tedious but straightforward exercise. In contrast, pCCD lacks a systematic correction scheme capable of seamlessly incorporating both weak and strong correlation.\cite{parkhill:2009,parkhill:2010a,boguslawski:2015,lehtola:2016,boguslawski:2017,lehtola:2018,boguslawski:2021,lehtola:2025}
	
	\acknowledgments{
	PAJ thanks the Natural Sciences and Engineering Research Council of Canada (Grant No. RGPIN-2024-05610) for funding. This research was made possible in part by the Digital Research Alliance of Canada.
	PFL thanks the European Research Council (ERC) under the European Union's Horizon 2020 research and innovation programme (Grant agreement No.~863481) for funding.}
	
	\section*{Data availability statement}
	The data that supports the findings of this study are available within the article.
	
	\section*{Conflict of interest}
	The authors have no conflicts of interest to disclose.

	\appendix
	\section{PP-EN2 expressions} \label{sec:pt2}
	\subsection{RDM elements}
		RDM elements for the intra-pair excitation $\ket{\PP^\alpha_\alpha}$ are almost the same as for $\ket{\PP}$. The only difference for the 1-RDM occurs in the VBS that has been modified
		\begin{equation} \label{eq:single_swap_1dm}
			(n_{\alpha_\mu})^\alpha_\alpha = \frac{1}{2}\qty[1 - \frac{(-1)^\mu \omega_\alpha}{\sqrt{\omega^2_\alpha +1}}]
			= n_{\alpha_{1-\mu}}
		\end{equation} 
		while the pair-transfer element in the modified VBS becomes
		\begin{equation} \label{eq:single_swap_pij}
			(P_{\ger{\alpha} \ung{\alpha}})^\alpha_\alpha 
			= (P_{\ung{\alpha} \ger{\alpha}})^\alpha_\alpha 
			= + \sqrt{(n_{\ger{\alpha}})^\alpha_\alpha (n_{\ung{\alpha}})^\alpha_\alpha}
			= - P_{\ger{\alpha} \ung{\alpha}}
		\end{equation}
		Symbolically, the density-density elements do not change at all, they remain the product of the 1-RDM elements (which are themselves modified) as in Eq.~\eqref{eq:GVB_2dm_D}.
	
		RDM elements for the inter-pair excitations $\ket{\PP^\beta_\alpha}$ are also simple modifications of those for $\ket{\PP}$. In particular, the $\alpha$th VBS is now empty, while the $\beta$th VBS is full. As a result, the 1-RDM elements are integers, i.e.,
		\begin{subequations}
		\begin{align} \label{eq:single_xfer_1dm}
			(n_{\ger{\alpha}})^\beta_\alpha = (n_{\ung{\alpha}})^\beta_\alpha &= 0 
			\\
			(n_{\ger{\beta}})^\beta_\alpha = (n_{\ung{\beta}})^\beta_\alpha &= 1
		\end{align}
		\end{subequations}
		while pair-transfer is not possible in \emph{neither} the empty nor the full VBS
		\begin{equation} \label{eq:single_xfer_pij}
			(P_{\ger{\alpha} \ung{\alpha}})^\beta_\alpha 
			= (P_{\ung{\alpha} \ger{\alpha}})^\beta_\alpha 
			= (P_{\ger{\beta} \ung{\beta}})^\beta_\alpha 
			= (P_{\ung{\beta} \ger{\beta}})^\beta_\alpha = 0.
		\end{equation}
		Again, the density-density elements are the product of the individual 1-RDM elements, as they are in Eq.~\eqref{eq:GVB_2dm_D}, with the additional element
		\begin{align}
			(D_{\ger{\beta} \ung{\beta}})^{\beta}_{\alpha} =1.
		\end{align}
		
		RDM elements for the double pair excitation $\ket*{\PP^{\alpha\beta}_{\alpha\beta}}$ are the same as Eqs.~\eqref{eq:single_swap_1dm} and \eqref{eq:single_swap_pij} but in \emph{both} the $\alpha$th and the $\beta$th VBS. Again, the density-density elements do not change symbolically.
		
	\subsection{Energy denominators}
		To compute the energy denominators, the primitive objects are those for the intra-pair excitations
		\begin{equation}
			\qty( \omega^2_\alpha + 1 )^{\frac{1}{2}}(E^{(0)} - E^{\alpha}_{\alpha}) = 
			- 2 L_{\ger{\alpha} \ung{\alpha}}
			+  2\omega_{\alpha} \sum_\mu (-1)^\mu 
			\qty(
			h_{\alpha_\mu \alpha_\mu} + \frac{1}{2} L_{\alpha_\mu \alpha_\mu}
			+ \sum_{\beta (\neq \alpha)} \sum_{\nu} G_{\alpha_\mu \beta_\nu} 
			n_{\beta_\nu} )
		\end{equation}
		which, because of the stationarity of the energy with respect to the energy gaps $\omega_{\alpha}$ [see Eq.~\eqref{eq:grad_vanish}], reduce to
		\begin{equation}
			E^{(0)} - E^{\alpha}_{\alpha} 
			= \frac{L_{\ger{\alpha} \ung{\alpha}}}{P_{\ger{\alpha}\ung{\alpha}}}.
		\end{equation}
		Energy denominators for double pair excitations are obtained as an update
		\begin{equation}
		\begin{split}
			E^{(0)} - E^{\alpha \beta}_{\alpha \beta} &= 
			\frac{L_{\ger{\alpha} \ung{\alpha}}}{P_{\ger{\alpha}\ung{\alpha}}}
			+ \frac{L_{\ger{\beta} \ung{\beta}}}{P_{\ger{\beta}\ung{\beta}}} 
			\\
			&-2 \qty(n_{\ung{\alpha}} - n_{\ger{\alpha}} ) \qty(n_{\ung{\beta}} - n_{\ger{\beta}} )
			\sum_{\mu\nu} (-1)^{\mu+\nu} G_{\alpha_\mu \beta_\nu}.
		\end{split}
		\end{equation}		
		With the electronic Hessian \eqref{eq:el_hess}, this may also be written
		\begin{equation}
		\begin{split}
			E^{(0)} - E^{\alpha \beta}_{\alpha \beta} &= 
			-2 \qty(\omega^2_{\alpha} + 1)^2 \pdv[2]{E}{\omega_\alpha}
			-2 \qty(\omega^2_{\beta} + 1)^2  \pdv[2]{E}{\omega_\beta}
			\\
			&- 4 \omega_{\alpha} \omega_{\beta} \qty(\omega^2_{\alpha} + 1) \qty(\omega^2_{\beta} + 1)
			\pdv[2]{E}{\omega_\alpha}{\omega_\beta}.
		\end{split}
		\end{equation}
		The energy denominators for the inter-pair excitations are likewise an update of the intra-pair excitations
		\begin{equation}
		\begin{split}
			E^{(0)} - E^{\beta}_{\alpha} &= 
			\frac{L_{\ger{\alpha} \ung{\alpha}}}{2P_{\ger{\alpha} \ung{\alpha}}}
			+ \frac{L_{\ger{\beta} \ung{\beta}}}{2P_{\ger{\beta} \ung{\beta}}}
			\\
			&+ \varepsilon_{\ger{\alpha}} + \varepsilon_{\ung{\alpha}} - \varepsilon_{\ger{\beta}} - \varepsilon_{\ung{\beta}}
			- 2 G_{\ger{\beta} \ung{\beta}} 
			\\
			&+ 2 \sum_{\mu\nu} G_{\alpha_\mu \beta_\nu} n_{\alpha_\mu} n_{\beta_{1-\nu}},
		\end{split}
		\end{equation}
		where $\varepsilon$ are the orbital energies defined in Eq.~\eqref{eq:pp_orb_energies}.
	
	\subsection{Transition elements}
		Energy numerators require transition density matrix (TDM) elements from the reference wavefunction $\ket{\PP}$ to its excited states. The intra-pair excitation $\ket{\PP^\alpha_\alpha}$ only couples to the reference through elements involving the $\alpha$th VBS. The normalized non-zero one-body elements are
		\begin{equation}
			\mel{\PP^\alpha_\alpha} {\n{\ger{\alpha}} } {\PP} 
			= -\mel{\PP^\alpha_\alpha} {\n{\ung{\alpha}} } {\PP} 
			= - P_{\ger{\alpha} \ung{\alpha}}
		\end{equation}
		which evidently sum to zero. There are also non-zero pair-transfer elements in the $\alpha$th VBS, which, when normalized, become
		\begin{subequations}
		\begin{align}
			\mel{\PP^\alpha_\alpha} {\Sp{\ung{\alpha}} \Sm{\ger{\alpha}} } {\PP} &= 
			 n_{\ger{\alpha}}
			\\
			\mel{\PP^\alpha_\alpha} {\Sp{\ger{\alpha}} \Sm{\ung{\alpha}} } {\PP} &= 
			 - n_{\ung{\alpha}}.
		\end{align}
		\end{subequations}
		Notice that pair-transfer TDM elements are \emph{not} symmetric. Non-zero density-density elements only occur when one index is in the $\alpha$th VBS and the other is not. When normalized, we have
		\begin{equation}
			\mel{\PP^\alpha_\alpha}  {\n{\ger{\alpha}} \n{\beta_\nu}}  {\PP} 
			= - \mel{\PP^\alpha_\alpha} { \n{\ung{\alpha}} \n{\beta_\nu}}  {\PP} 
			= \mel{\PP^\alpha_\alpha}  {\n{\ger{\alpha}}}  {\PP} n_{\beta_\nu}
		\end{equation}
		where, in the right-hand side, the 1-RDM elements $(n_{\beta_\nu})^\alpha_\alpha = n_{\beta_\nu}$ are identical as $\beta_\nu$ is not in the $\alpha$th VBS.
		
		The double pair excitation $\ket*{\PP^{\alpha\beta}_{\alpha\beta}}$ couples to the reference $\ket{\PP}$ \emph{only} through the normalized elements
		\begin{equation}
		\begin{split}
			  \mel*{\PP^{\alpha\beta}_{\alpha\beta}} {\n{\ger{\alpha}} \n{\ger{\beta}}}  {\PP} &= 
			- \mel*{\PP^{\alpha\beta}_{\alpha\beta}} {\n{\ger{\alpha}} \n{\ung{\beta}}}  {\PP} =
			- \mel*{\PP^{\alpha\beta}_{\alpha\beta}} {\n{\ung{\alpha}} \n{\ger{\beta}}}  {\PP} =
			  \mel*{\PP^{\alpha\beta}_{\alpha\beta}} {\n{\ung{\alpha}} \n{\ung{\beta}}}  {\PP}  
			  \\
			&= P_{\ger{\alpha}\ung{\alpha}} P_{\ger{\beta}\ung{\beta}}
		\end{split}
		\end{equation}
		Any state with more than two pair excitations will not couple to the reference $\ket{\PP}$ through a two-electron operator.
		
		The inter-pair excitation $\ket{\PP^{\beta}_{\alpha}}$ only couples to $\ket{\PP}$ through pair-transfers taking a pair of electrons from the $\alpha$th VBS and placing it in the $\beta$th. The non-zero normalized elements are
		\begin{align}
			\mel*{\PP^{\beta}_{\alpha}} {\Sp{\beta_{1-\nu}} \Sm{\alpha_{\mu}}} {\PP} &= 
			(-1)^{\mu+\nu} \sqrt{n_{\alpha_{\mu}} n_{\beta_{\nu}} }.
		\end{align}
		Again, pair-transfer TDM elements are not symmetric, and they are forbidden in the reverse processes
		\begin{align}
			  \mel*{\PP^{\beta}_{\alpha}} {\Sp{\ger{\alpha}} \Sm{\ger{\beta}}} {\PP} 
			= \mel*{\PP^{\beta}_{\alpha}} {\Sp{\ung{\alpha}} \Sm{\ger{\beta}}} {\PP} 
			= \mel*{\PP^{\beta}_{\alpha}} {\Sp{\ger{\alpha}} \Sm{\ung{\beta}}} {\PP} 
			= \mel*{\PP^{\beta}_{\alpha}} {\Sp{\ung{\alpha}} \Sm{\ung{\beta}}} {\PP} = 0.
		\end{align}
		
		The energy numerators are thus simple to evaluate, giving
		\begin{align}
			\qty( \omega^2_\alpha + 1 )^{\frac{1}{2}} \mel{\PP^{\alpha}_{\alpha}}{\hH_\text{C}}{\PP} =
			\omega_{\alpha} L_{\ger{\alpha} \ung{\alpha}}
			+ \sum_\mu (-1)^\mu
			\qty(
			h_{\alpha_\mu \alpha_\mu} + \frac{1}{2}L_{\alpha_\mu \alpha_\mu} + \sum_{\beta (\neq \alpha)} \sum_\nu G_{\alpha_\mu \beta_\nu} n_{\beta_\nu}
			)
		\end{align}
		which \emph{vanishes} as a consequence of the stationarity of the energy given by Eq.~\eqref{eq:grad_vanish}. The intra-pair transfers thus give no contribution, analogous to the Brillouin theorem for Slater determinants in the HF orbitals. The energy numerators for the double pair excitations are 
		\begin{equation}
			\mel*{\PP^{\alpha\beta}_{\alpha\beta}}{\hH_\text{C}}{\PP}
			= 2 P_{\ger{\alpha}\ung{\alpha}} P_{\ger{\beta}\ung{\beta}} \sum_{\mu\nu} (-1)^{\mu+\nu} G_{\alpha_\mu \beta_\nu}
		\end{equation}
		or, from Eq.~\eqref{eq:el_hess}, we have
		\begin{equation}
			\mel*{\PP^{\alpha\beta}_{\alpha\beta}}{\hH_\text{C}}{\PP} 
			= 2 (\omega^2_{\alpha}+1)(\omega^2_{\beta}+1) \pdv[2]{E}{\omega_\alpha}{\omega_\beta}.
		\end{equation}
		For the inter-pair excitations, the energy numerators are likewise simple
		\begin{equation}
			\begin{split}
				\mel{\PP^{\beta}_{\alpha}}{\hH_\text{C}} {\PP} = 
				  L_{\beta_1 \alpha_0} \sqrt{n_{\ger{\alpha}} n_{\ger{\beta}} }
				- L_{\beta_1 \alpha_1} \sqrt{n_{\ung{\alpha}} n_{\ger{\beta}} }
				- L_{\beta_0 \alpha_0} \sqrt{n_{\ger{\alpha}} n_{\ung{\beta}} }
				+ L_{\beta_0 \alpha_1} \sqrt{n_{\ung{\alpha}} n_{\ung{\beta}} }.
			\end{split}
		\end{equation}

\bibliography{pCCD_GVB}

@article{jerzy:2024,
	author = {Cioslowski, J. and Strasburger, K.},
	title = {Constraints upon Functionals of the 1-Matrix, Universal Properties of Natural Orbitals, and the Fallacy of the Collins {``}Conjecture{''}},
	journal = {J. Phys. Chem. Lett.},
	volume = {15},
	number = {5},
	pages = {1328-1337},
	doi = {10.1021/acs.jpclett.3c03118},
	year = {2024}
}

@article{zamani:2025,
	author = {Zamani, A. Y. and Carter-Fenk, K.},
	title = {Toward ab initio relatizations of {C}ollins's conjecture},
	journal = {J. Chem. Phys.},
	volume = {163},
	number = {3},
	pages = {034103},
	doi = {10.1063/5.0274457},
	year = {2025}
}

@article{bloch:1965,
	author = {Bloch, F.},
	title = {Off-Diagonal Long-Range Order and Persistent Currents in a Hollow Cylinder},
	journal = {Phys. Rev.},
	volume = {137},
	number = {3A},
	pages = {A787-A795},
	doi = {10.1103/PhysRev.137.A787},
	year = {1965}
}

@article{jaynes:1957a,
	author = {Jaynes, E. T.},
	title = {Information Theory and Statistical Mechanics},
	journal = {Phys. Rev.},
	volume = {106},
	number = {4},
	pages = {620-630},
	doi = {10.1103/PhysRev.106.620},
	year = {1957}
}

@article{jaynes:1957b,
	author = {Jaynes, E. T.},
	title = {Information Theory and Statistical Mechanics. {II.}},
	journal = {Phys. Rev.},
	volume = {108},
	number = {2},
	pages = {171-190},
	doi = {10.1103/PhysRev.108.171},
	year = {1957}
}

@article{collins:1993,
	author = {Collins, D. M.},
	title = {Entropy Maximizations on Electron Density},
	journal = {Z. Naturforsch. A},
	pages = {68-74},
	volume = {48},
	number = {1-2},
	doi = {10.1515/zna-1993-1-218},
	year = {1993}
}

@article{hait:2023,
	author = {Hait, D. and Head-Gordon, M.},
	journal = {Angew. Chem.},
	title = {When is a bond broken? {T}he polarizability perspective},
	doi = {10.1002/ange.202312078},
	volume = {135},
	number = {46},
	pages = {e202312078},
	year = {2023}
}

@article{tian:2001,
	author = {Tian, G.-S. and Tang, L.-H. and Chen, Q.-H.},
	title = {Pair-mixing superconducting correlation in ultrasmall metallic grains},
	journal = {Phys. Rev. B},
	volume = {63},
	number = {5},
	pages = {054511},
	doi = {10.1103/PhysRevB.63.054511},
	year = {2001}
}

@article{penrose:1956,
	author = {Penrose, O. and Onsager, L.},
	title = {Bose-{E}instein condensation and Liquid Helium},
	journal = {Phys. Rev.},
	volume = {104},
	number = {3},
	pages = {576-584},
	doi = {10.1103/PhysRev.104.576},
	year = {1956}
}

@article{werner:1985,
	author = {Werner, Hans-Joachim and Knowles, Peter J.},
	doi = {10.1063/1.448627},
	journal = {J. Chem. Phys.},
	number = {11},
	pages = {5053-5063},
	title = {A second order multiconfiguration {SCF} procedure with optimum convergence},
	volume = {82},
	year = {1985}}

@article{liebenthal:2025,
	author = {Liebenthal, Marcus D. and Yuwono, Stephen H. and Koulias, Lauren N. and Li, Run R. and Rubin, Nicholas C. and DePrince III, A. Eugene},
	doi = {10.1021/acs.jpca.5c00329},
	journal = {J. Phys. Chem. A},
	number = {29},
	pages = {6679-6693},
	title = {Automated Quantum Chemistry Code Generation with the p†q Package},
	volume = {129},
	year = {2025}}

@article{Frisch_1990,
	author = {Michael J. Frisch and Martin Head-Gordon and John A. Pople},
	doi = {https://doi.org/10.1016/0009-2614(90)80030-H},
	journal = {Chem. Phys. Lett.},
	number = {3},
	pages = {281-289},
	title = {Semi-direct algorithms for the MP2 energy and gradient},
	volume = {166},
	year = {1990}}

@article{mitxelena:2018,
	author = {Mitxelena, I. and Rodr\'{i}guez-Mayorga, M. and Piris, M.},
	doi = {10.1140/epjb/e2018-90078-8},
	journal = {Eur. Phys. J. B},
	pages = {109},
	title = {Phase dilemma in natural orbital functional theory from the {N}-representability perspective},
	volume = {91},
	year = {2018}}

@article{hapka:2022,
	author = {Hapka, M. and Pernal, K. and Jensen, H. J. A.},
	doi = {10.1063/5.0082155},
	journal = {J. Chem. Phys.},
	number = {17},
	pages = {174102},
	title = {An efficient implementation of time-dependent linear-response theory for strongly orthogonal geminal wave function models},
	volume = {156},
	year = {2022}}

@article{johnson:2025b,
	author = {Johnson, Paul A.},
	doi = {10.1063/5.0257950},
	journal = {J. Chem. Phys.},
	number = {13},
	pages = {134106},
	title = {Richardson-{G}audin states of non-zero seniority: {M}atrix elements},
	volume = {162},
	year = {2025}}

@article{johnson:2025c,
	author = {Johnson, Paul A.},
	doi = {10.1063/5.0284864},
	journal = {J. Chem. Phys.},
	number = {8},
	pages = {081101},
	title = {Richardson-{G}audin states of non-zero seniority. {II}. {S}ingle-reference treatment of strong correlation},
	volume = {163},
	year = {2025}}

@article{piris2021donof,
	author = {Piris, Mario and Mitxelena, Ion},
	doi = {10.1016/j.cpc.2020.107651},
	journal = {Comp. Phys. Comm.},
	pages = {107651},
	publisher = {Elsevier},
	title = {{DoNOF}: An open-source implementation of natural-orbital-functional-based methods for quantum chemistry},
	volume = {259},
	year = {2021}}

@article{bruneval2016molgw,
	author = {Bruneval, F. and Rangel, T. and Hamed, S.M. and Shao, M. and Yang, C. and Neaton, J.B.},
	doi = {10.1016/j.cpc.2016.06.019},
	journal = {Comp. Phys. Comm.},
	pages = {149},
	title = {molgw 1: Many-body perturbation theory software for atoms, molecules, and clusters},
	volume = {208},
	year = {2016}}

@misc{rodriguez2022snoft,
	author = {Rodr\'iguez-Mayorga, M.},
	journal = {Zenodo},
	publisher = {ACS Publications},
	title = {Standalone {NOFT} module (1.0) Published on {Z}enodo},
	year = {2022}}

@article{rodriguez-mayorga:2025,
	author = {Rodr{\'\i}guez-Mayorga, Mauricio and Loos, Pierre-Fran{\c c}ois and Bruneval, Fabien and Visscher, Lucas},
	doi = {10.1063/5.0242504},
	journal = {J. Chem. Phys.},
	month = {02},
	number = {5},
	pages = {054716},
	title = {Time-reversal symmetry in RDMFT and pCCD with complex-valued orbitals},
	volume = {162},
	year = {2025}}

@article{jahani:2025,
	author = {Jahani, S. and Ahmadkhani, S. and Boguslawski, K. and Tecmer, P.},
	doi = {10.1063/5.0262453},
	journal = {J. Chem. Phys.},
	number = {18},
	pages = {184110},
	title = {Simple and efficient computational strategies for calculating orbital energies and pair-orbital energies from {pCCD}-based methods},
	volume = {162},
	year = {2025}}

@article{yao:2021,
	author = {Yao, Y. and Umrigar, C. J.},
	doi = {10.1021/acs.jctc.1c00385},
	journal = {J. Chem. Theory Comput.},
	number = {7},
	pages = {4183-4194},
	title = {Orbital Optimization in Selected Configuration Interaction Methods},
	volume = {17},
	year = {2021}}

@article{sun:2018,
	author = {Sun, Qiming and Berkelbach, Timothy C. and Blunt, Nick S. and Chan, Garnet Kin-Lic and Zheng, Bo-Xiao and Guo, Dan and Li, Zhendong},
	doi = {10.1002/wcms.1340},
	journal = {WIREs Comput. Mol. Sci.},
	number = {1},
	pages = {e1340},
	title = {{PySCF: the Python-based simulations of chemistry framework}},
	volume = {8},
	year = {2018}}

@article{kottmann:2024,
	author = {Kottmann, Jakob S. and Scala, Francesco},
	doi = {10.1021/acs.jctc.3c00565},
	journal = {J. Chem. Theory Comput.},
	number = {9},
	pages = {3514-3523},
	title = {Quantum Algorithmic Approach to Multiconfigurational Valence Bond Theory: Insights from Interpretable Circuit Design},
	volume = {20},
	year = {2024}}

@article{kottmann:2023,
	author = {Kottmann, Jakob S.},
	doi = {10.22331/q-2023-08-03-1073},
	journal = {{Quantum}},
	pages = {1073},
	title = {Molecular {Q}uantum {C}ircuit {D}esign: {A} {G}raph-{B}ased {A}pproach},
	volume = {7},
	year = {2023}}

@article{larsson:2020,
	author = {Larsson, Henrik R. and Jim{\'e}nez-Hoyos, Carlos A. and Chan, Garnet Kin-Lic},
	doi = {10.1021/acs.jctc.0c00463},
	journal = {J. Chem. Theory Comput.},
	number = {8},
	pages = {5057-5066},
	title = {Minimal Matrix Product States and Generalizations of Mean-Field and Geminal Wave Functions},
	volume = {16},
	year = {2020}}

@article{kottmann:2022,
	author = {Kottmann, Jakob S. and Aspuru-Guzik, Al\'an},
	doi = {10.1103/PhysRevA.105.032449},
	journal = {Phys. Rev. A},
	pages = {032449},
	title = {Optimized low-depth quantum circuits for molecular electronic structure using a separable-pair approximation},
	volume = {105},
	year = {2022}}

@article{burton:2024a,
	author = {Burton, Hugh G. A.},
	doi = {10.1103/PhysRevResearch.6.023300},
	issue = {2},
	journal = {Phys. Rev. Res.},
	month = {Jun},
	numpages = {17},
	pages = {023300},
	publisher = {American Physical Society},
	title = {Accurate and gate-efficient quantum Ans\"atze for electronic states without adaptive optimization},
	url = {https://link.aps.org/doi/10.1103/PhysRevResearch.6.023300},
	volume = {6},
	year = {2024}}

@article{thorsteinsson:1997,
	author = {Thorsteinsson, Thorstein and Cooper, David L. and Gerratt, Joseph and Raimondi, Mario},
	doi = {10.1007/bf02341697},
	journal = {Theor. Chem. Acc.},
	number = {3--4},
	pages = {131--150},
	title = {Symmetry adaptation and the utilization of point group symmetry in valence bond calculations, including {CASVB}},
	volume = {95},
	year = {1997}}

@article{thorsteinsson:1996b,
	author = {Thorsteinsson, Thorstein and Cooper, David L.},
	doi = {10.1007/bf00186445},
	journal = {Theor. Chem. Acc.},
	number = {4},
	pages = {233--245},
	title = {Exact transformations of {CI} spaces, {VB} representations of {CASSCF} wavefunctions and the optimization of {VB} wavefunctions},
	volume = {94},
	year = {1996}}

@article{thorsteinsson:1996a,
	author = {Thorsteinsson, Thorstein and Cooper, David L. and Gerratt, Joseph and Karadakov, Peter B. and Raimondi, Mario},
	doi = {10.1007/bf01129215},
	journal = {Theor. Chem. Acc.},
	number = {6},
	pages = {343--366},
	title = {Modern valence bond representations of {CASSCF} wavefunctions},
	volume = {93},
	year = {1996}}

@article{dykstra:1980,
	author = {Dykstra, Clifford E.},
	doi = {10.1063/1.439492},
	journal = {J. Chem. Phys.},
	month = {03},
	number = {5},
	pages = {2928-2935},
	title = {Perfect pairing valence bond generalization of self‐consistent electron pair theory},
	volume = {72},
	year = {1980}}

@article{parkhill:2010a,
	author = {Parkhill, John A. and Head-Gordon, Martin},
	doi = {10.1063/1.3456001},
	journal = {J. Chem. Phys.},
	month = {07},
	number = {2},
	pages = {024103},
	title = {A tractable and accurate electronic structure method for static correlations: The perfect hextuples model},
	volume = {133},
	year = {2010}}

@article{burton:2024b,
	author = {Burton, Hugh G. A.},
	doi = {10.1039/D4FD00064A},
	issue = {0},
	journal = {Faraday Discuss.},
	pages = {157-169},
	publisher = {The Royal Society of Chemistry},
	title = {Tiled unitary product states for strongly correlated Hamiltonians},
	url = {http://dx.doi.org/10.1039/D4FD00064A},
	volume = {254},
	year = {2024}}

@article{parkhill:2010b,
	author = {Parkhill, John A. and Head-Gordon, Martin},
	doi = {10.1063/1.3483556},
	journal = {J. Chem. Phys.},
	month = {09},
	number = {12},
	pages = {124102},
	title = {A truncation hierarchy of coupled cluster models of strongly correlated systems based on perfect-pairing references: The singles+doubles models},
	volume = {133},
	year = {2010}}

@article{beran:2006,
	author = {Beran, Gregory J. O. and Head-Gordon, Martin and Gwaltney, Steven R.},
	doi = {10.1063/1.2176603},
	journal = {J. Chem. Phys.},
	month = {03},
	number = {11},
	pages = {114107},
	title = {Second-order correction to perfect pairing: An inexpensive electronic structure method for the treatment of strong electron-electron correlations},
	volume = {124},
	year = {2006}}

@article{beran:2005,
	author = {Beran, Gregory J. O. and Austin, Brian and Sodt, Alex and Head-Gordon, Martin},
	doi = {10.1021/jp053780c},
	journal = {J. Phys. Chem. A},
	number = {40},
	pages = {9183-9192},
	title = {Unrestricted Perfect Pairing: The Simplest Wave-Function-Based Model Chemistry beyond Mean Field},
	volume = {109},
	year = {2005}}

@article{lehtola:2016,
	author = {Lehtola, Susi and Parkhill, John and Head-Gordon, Martin},
	doi = {10.1063/1.4964317},
	journal = {J. Chem. Phys.},
	month = {10},
	number = {13},
	pages = {134110},
	title = {Cost-effective description of strong correlation: Efficient implementations of the perfect quadruples and perfect hextuples models},
	volume = {145},
	year = {2016}}

@article{lehtola:2018,
	author = {Lehtola, S. and Parkhill, J. and Head-Gordon, M.},
	doi = {10.1080/00268976.2017.1342009},
	journal = {Mol. Phys.},
	number = {5-6},
	pages = {547--560},
	title = {Orbital optimisation in the perfect pairing hierarchy: applications to full-valence calculations on linear polyacenes},
	volume = {116},
	year = {2018}}

@article{lehtola:2025,
	author = {Lehtola, S. and Head-Gordon, M.},
	doi = {10.1080/00268976.2025.2489096},
	journal = {Mol. Phys.},
	pages = {e2489096},
	title = {Coupled-cluster pairing models for radicals with strong correlations},
	year = {2025}}

@article{lawler:2010,
	author = {Lawler, Keith V. and Small, David W. and Head-Gordon, Martin},
	doi = {10.1021/jp911009f},
	journal = {J. Phys. Chem. A},
	number = {8},
	pages = {2930-2938},
	title = {Orbitals That Are Unrestricted in Active Pairs for Generalized Valence Bond Coupled Cluster Methods},
	volume = {114},
	year = {2010}}

@article{small:2011,
	author = {Small, David W. and Head-Gordon, Martin},
	doi = {10.1039/C1CP21832H},
	journal = {Phys. Chem. Chem. Phys.},
	pages = {19285-19297},
	title = {Post-modern valence bond theory for strongly correlated electron spins},
	volume = {13},
	year = {2011}}

@article{small:2012,
	author = {Small, David W. and Head-Gordon, Martin},
	doi = {10.1063/1.4751485},
	journal = {J. Chem. Phys.},
	month = {09},
	number = {11},
	pages = {114103},
	title = {A fusion of the closed-shell coupled cluster singles and doubles method and valence-bond theory for bond breaking},
	volume = {137},
	year = {2012}}

@article{lewis:1916,
	author = {Lewis, Gilbert N.},
	doi = {10.1021/ja02261a002},
	journal = {J. Am. Chem. Soc.},
	number = {4},
	pages = {762-785},
	title = {The Atom And The Molecule},
	volume = {38},
	year = {1916}}

@article{dunning:2016,
	author = {Dunning, Thom H. Jr. and Xu, Lu T. and Takeshita, Tyler Y. and Lindquist, Beth A.},
	doi = {10.1021/acs.jpca.5b12335},
	journal = {J. Phys. Chem. A},
	number = {11},
	pages = {1763-1778},
	title = {Insights into the Electronic Structure of Molecules from Generalized Valence Bond Theory},
	volume = {120},
	year = {2016}}

@inbook{bobrowicz:1977,
	address = {Boston, MA},
	author = {Bobrowicz, Frank W. and Goddard, William A.},
	booktitle = {Methods of Electronic Structure Theory},
	doi = {10.1007/978-1-4757-0887-5_4},
	editor = {Schaefer, Henry F.},
	isbn = {978-1-4757-0887-5},
	pages = {79--127},
	publisher = {Springer US},
	title = {The Self-Consistent Field Equations for Generalized Valence Bond and Open-Shell {H}artree--{F}ock Wave Functions},
	url = {https://doi.org/10.1007/978-1-4757-0887-5_4},
	year = {1977}}

@book{szabo_book,
	address = {New York},
	author = {A. Szabo and N. S. Ostlund},
	publisher = {McGraw-Hill},
	title = {Modern quantum chemistry},
	year = {1989}}

@book{peierls_book,
	address = {London},
	author = {R. Peierls},
	publisher = {Oxford University Press},
	title = {Quantum Theory of Solids},
	year = {1955}}

@book{jorgersen_book,
	author = {P. J{\o}rgensen and J. Simons},
	doi = {10.1016/b978-0-12-390220-7.x5001-6},
	isbn = {9780123902207},
	publisher = {Elsevier},
	title = {Second Quantization-Based Methods in Quantum Chemistry},
	url = {http://dx.doi.org/10.1016/B978-0-12-390220-7.X5001-6},
	year = {1981}}

@article{cooper:1956,
	author = {Cooper, Leon N.},
	doi = {10.1103/PhysRev.104.1189},
	issue = {4},
	journal = {Phys. Rev.},
	month = {Nov},
	numpages = {0},
	pages = {1189--1190},
	publisher = {American Physical Society},
	title = {Bound Electron Pairs in a Degenerate {F}ermi Gas},
	url = {https://link.aps.org/doi/10.1103/PhysRev.104.1189},
	volume = {104},
	year = {1956}}

@article{wouters:2014,
	author = {Wouters, Sebastian and Van Neck, Dimitri},
	doi = {10.1140/epjd/e2014-50500-1},
	journal = {Eur. Phys. J. D},
	number = {9},
	pages = {272},
	title = {The density matrix renormalization group for ab initio quantum chemistry},
	volume = {68},
	year = {2014}}

@article{szalay:2015,
	author = {Szalay, Szil{\'a}rd and Pfeffer, Max and Murg, Valentin and Barcza, Gergely and Verstraete, Frank and Schneider, Reinhold and Legeza, {\"O}rs},
	doi = {https://doi.org/10.1002/qua.24898},
	journal = {Int. J. Quantum Chem.},
	number = {19},
	pages = {1342-1391},
	title = {Tensor product methods and entanglement optimization for ab initio quantum chemistry},
	volume = {115},
	year = {2015}}

@incollection{johnson:2024a,
	author = {Paul A. Johnson},
	booktitle = {Novel Treatments of Strong Correlations},
	doi = {https://doi.org/10.1016/bs.aiq.2024.04.003},
	editor = {Ramon A. Miranda Quintana and John F. Stanton},
	issn = {0065-3276},
	pages = {67-119},
	publisher = {Academic Press},
	series = {Advances in Quantum Chemistry},
	title = {{Richardson--Gaudin states}},
	url = {https://www.sciencedirect.com/science/article/pii/S0065327624000170},
	volume = {90},
	year = {2024}}

@incollection{debaerdemacker:2024,
	author = {Stijn {De Baerdemacker} and Dimitri {Van Neck}},
	booktitle = {Novel Treatments of Strong Correlations},
	doi = {https://doi.org/10.1016/bs.aiq.2024.07.002},
	editor = {Ramon A. Miranda Quintana and John F. Stanton},
	issn = {0065-3276},
	pages = {185-218},
	publisher = {Academic Press},
	series = {Advances in Quantum Chemistry},
	title = {Geminal theory within the seniority formalism and bi-variational principle},
	url = {https://www.sciencedirect.com/science/article/pii/S0065327624000200},
	volume = {90},
	year = {2024}}

@article{alcoba:2014b,
	author = {Alcoba, Diego R. and Torre, Alicia and Lain, Luis and O{\~n}a, Ofelia B. and Capuzzi, Pablo and Van Raemdonck, Mario and Bultinck, Patrick and Van Neck, Dimitri},
	doi = {10.1063/1.4904755},
	journal = {J. Chem. Phys.},
	month = {12},
	number = {24},
	pages = {244118},
	title = {A hybrid configuration interaction treatment based on seniority number and excitation schemes},
	volume = {141},
	year = {2014}}

@article{smith:1965,
	author = {Smith, Darwin W. and Fogel, Sidney J.},
	doi = {10.1063/1.1701519},
	journal = {J. Chem. Phys.},
	month = {11},
	number = {10},
	pages = {S91-S96},
	title = {Natural Orbitals and Geminals of the Beryllium Atom},
	volume = {43},
	year = {1965}}

@article{kossoski:2023,
	author = {Kossoski, F{\'a}bris and Loos, Pierre-Fran{\c{c}}ois},
	doi = {10.1021/acs.jctc.3c00946},
	journal = {J. Chem. Theory Comput.},
	number = {23},
	pages = {8654-8670},
	title = {Seniority and Hierarchy Configuration Interaction for Radicals and Excited States},
	volume = {19},
	year = {2023}}

@article{coleman:1963,
	author = {Coleman, A. J.},
	doi = {10.1103/RevModPhys.35.668},
	issue = {3},
	journal = {Rev. Mod. Phys.},
	month = {Jul},
	numpages = {0},
	pages = {668--686},
	publisher = {American Physical Society},
	title = {Structure of Fermion Density Matrices},
	url = {https://link.aps.org/doi/10.1103/RevModPhys.35.668},
	volume = {35},
	year = {1963}}

@article{allen:1961,
	author = {Allen, Thomas L. and Shull, Harrison},
	doi = {10.1063/1.1732124},
	journal = {J. Chem. Phys.},
	month = {11},
	number = {5},
	pages = {1644-1651},
	title = {The Chemical Bond in Molecular Quantum Mechanics},
	volume = {35},
	year = {1961}}

@article{shull:1959,
	author = {Shull, Harrison},
	doi = {10.1063/1.1730212},
	journal = {J. Chem. Phys.},
	month = {06},
	number = {6},
	pages = {1405-1413},
	title = {Natural Spin Orbital Analysis of Hydrogen Molecule Wave Functions},
	volume = {30},
	year = {1959}}

@article{kucharski:1992,
	author = {Kucharski, S. A. and Bartlett, R. J.},
	doi = {10.1063/1.463930},
	journal = {J. Chem. Phys.},
	pages = {4282},
	title = {The Coupled-Cluster Single, Double, Triple, and Quadruple Excitation Method},
	volume = {97},
	year = {1992}}

@article{oliphant:1991,
	author = {Oliphant, N. and Adamowicz, L.},
	doi = {10.1063/1.461534},
	journal = {J. Chem. Phys.},
	pages = {6645},
	title = {Coupled-Cluster Method Truncated at Quadruples},
	volume = {95},
	year = {1991}}

@article{giner:2013,
	author = {Giner, Emmanuel and Scemama, Anthony and Caffarel, Michel},
	doi = {10.1139/cjc-2013-0017},
	journal = {Can. J. Chem.},
	number = {9},
	pages = {879},
	title = {{Using perturbatively selected configuration interaction in quantum Monte Carlo calculations}},
	volume = {91},
	year = {2013}}

@article{huron:1973,
	author = {Huron, B. and Malrieu, J. P. and Rancurel, P.},
	doi = {10.1063/1.1679199},
	journal = {J. Chem. Phys.},
	pages = {5745},
	title = {{Iterative perturbation calculations of ground and excited state energies from multiconfigurational zeroth-order wave functions}},
	volume = {58},
	year = {1973}}

@article{liu:2014,
	author = {Xinle Liu and Joseph E. Subotnik},
	doi = {10.1021/ct4009377},
	journal = {J. Chem. Theory Comput.},
	pages = {1004},
	title = {{The Variationally Orbital-Adapted Configuration Interaction Singles (VOA-CIS) Approach to Electronically Excited States}},
	volume = {10},
	year = {2014}}

@article{holmes:2016,
	author = {Holmes, Adam A. and Tubman, Norm M. and Umrigar, C. J.},
	doi = {10.1021/acs.jctc.6b00407},
	journal = {J. Chem. Theory Comput.},
	number = {8},
	pages = {3674},
	title = {{Heat-Bath Configuration Interaction: An Efficient Selected Configuration Interaction Algorithm Inspired by Heat-Bath Sampling}},
	volume = {12},
	year = {2016}}

@article{schriber:2016,
	author = {Schriber, Jeffrey B. and Evangelista, Francesco A.},
	doi = {10.1063/1.4948308},
	journal = {J. Chem. Phys.},
	number = {16},
	pages = {161106},
	title = {Communication: {An} adaptive configuration interaction approach for strongly correlated electrons with tunable accuracy},
	volume = {144},
	year = {2016}}

@article{garniron:2018,
	author = {Y. Garniron and A. Scemama and E. Giner and M. Caffarel and P. F. Loos},
	doi = {10.1063/1.5044503},
	journal = {J. Chem. Phys.},
	pages = {064103},
	title = {Selected Configuration Interaction Dressed by Perturbation},
	volume = {149},
	year = {2018}}

@article{baiardi:2020,
	author = {Baiardi, Alberto and Reiher, Markus},
	doi = {10.1063/1.5129672},
	journal = {J. Chem. Phys.},
	number = {4},
	pages = {040903},
	publisher = {AIP Publishing},
	title = {The density matrix renormalization group in chemistry and molecular physics: Recent developments and new challenges},
	volume = {152},
	year = {2020}}

@article{chan:2011,
	author = {Chan, Garnet Kin-Lic and Sharma, Sandeep},
	doi = {10.1146/annurev-physchem-032210-103338},
	journal = {Ann. Rev. Phys. Chem.},
	number = {1},
	pages = {465--481},
	publisher = {Annual Reviews},
	title = {The density matrix renormalization group in quantum chemistry},
	volume = {62},
	year = {2011}}

@article{white:1993,
	author = {White, Steven R},
	doi = {10.1103/PhysRevB.48.10345},
	journal = {Phys. Rev. B},
	number = {14},
	pages = {10345},
	publisher = {APS},
	title = {Density-matrix algorithms for quantum renormalization groups},
	volume = {48},
	year = {1993}}

@article{white:1992,
	author = {White, Steven R},
	doi = {10.1103/PhysRevLett.69.2863},
	journal = {Phys. Rev. Lett.},
	number = {19},
	pages = {2863},
	publisher = {APS},
	title = {Density matrix formulation for quantum renormalization groups},
	volume = {69},
	year = {1992}}

@incollection{roos:2005,
	author = {Roos, Bj{\"o}rn O},
	booktitle = {Theory and Applications of Computational Chemistry},
	doi = {10.1016/B978-044451719-7/50068-8},
	pages = {725--764},
	publisher = {Elsevier},
	title = {Multiconfigurational quantum chemistry},
	year = {2005}}

@article{roos:1980b,
	author = {Roos, Bj{\"o}rn O},
	doi = {10.1002/qua.560180822},
	journal = {Int. J. Quantum Chem.},
	number = {S14},
	pages = {175--189},
	publisher = {Wiley Online Library},
	title = {The complete active space {SCF} method in a {F}ock-matrix-based super-{CI} formulation},
	volume = {18},
	year = {1980}}

@article{roos:1980a,
	author = {Roos, Bj{\"o}rn O and Taylor, Peter R and Sigbahn, Per EM},
	doi = {10.1016/0301-0104(80)80045-0},
	journal = {Chem. Phys.},
	number = {2},
	pages = {157},
	publisher = {Elsevier},
	title = {A complete active space \text{SCF} method (\text{CASSCF}) using a density matrix formulated super-{CI} approach},
	volume = {48},
	year = {1980}}

@article{hirata:2003,
	author = {Hirata, S},
	doi = {10.1021/jp034596z},
	journal = {J. Phys. Chem. A},
	number = {46},
	pages = {9887-9897},
	title = {Tensor Contraction Engine: Abstraction and Automated Parallel Implementation of Configuration-Interaction, Coupled-Cluster, and Many-Body Perturbation Theories},
	volume = {107},
	year = {2003}}

@article{dirac:1927,
	author = {Dirac, Paul Adrien Maurice and Bohr, Niels Henrik David},
	doi = {10.1098/rspa.1927.0039},
	journal = {Proceedings of the Royal Society of London. Series A, Containing Papers of a Mathematical and Physical Character},
	number = {767},
	pages = {243-265},
	title = {The quantum theory of the emission and absorption of radiation},
	volume = {114},
	year = {1927}}

@article{quintero-monsebaiz:2023,
	author = {Quintero-Monsebaiz, Ra{\'u}l and Loos, Pierre-Fran{\c c}ois},
	doi = {10.1063/5.0163846},
	journal = {AIP Advances},
	number = {8},
	pages = {085035},
	title = {Equation generator for equation-of-motion coupled cluster assisted by computer algebra system},
	volume = {13},
	year = {2023}}

@incollection{crawford:2000,
	address = {Chichester, England, UK},
	author = {Crawford, T. Daniel and Schaefer, Henry F.},
	booktitle = {{Reviews in Computational Chemistry}},
	doi = {10.1002/9780470125915.ch2},
	isbn = {978-0-47012591-5},
	journal = {Wiley Online Library},
	month = jan,
	pages = {33--136},
	publisher = {John Wiley {\&} Sons, Ltd},
	title = {{An Introduction to Coupled Cluster Theory for Computational Chemists}},
	year = {2000}}

@article{rubin:2011,
	author = {Rubin, Nicholas C. and DePrince III, A. Eugene},
	doi = {10.1080/00268976.2021.1954709},
	issn = {0026-8976},
	journal = {Mol. Phys.},
	number = {21-22},
	pages = {e1954709},
	publisher = {Taylor {\&} Francis},
	title = {{{p$^\dagger$q}: a tool for prototyping many-body methods for quantum chemistry}},
	volume = {119},
	year = {2021}}

@article{evangelista:2022,
	author = {Evangelista, Francesco A.},
	doi = {10.1063/5.0097858},
	journal = {J. Chem. Phys.},
	number = {6},
	pages = {064111},
	title = {{Automatic derivation of many-body theories based on general Fermi vacua}},
	volume = {157},
	year = {2022}}

@article{wick:1950,
	author = {Wick, G. C.},
	doi = {10.1103/PhysRev.80.268},
	issn = {1536-6065},
	journal = {Phys. Rev.},
	month = oct,
	number = {2},
	pages = {268--272},
	publisher = {American Physical Society},
	title = {{The Evaluation of the Collision Matrix}},
	volume = {80},
	year = {1950}}

@article{alcoba:2014a,
	author = {Alcoba, D. R. and Torre, A. and Lain, L. and Massaccesi, G. E. and O\~{n}a, O.},
	doi = {10.1063/1.4882881},
	journal = {J. Chem. Phys.},
	number = {23},
	pages = {234103},
	title = {Configuration interaction wave functions: A seniority number approach},
	volume = {140},
	year = {2014}}

@article{bardeen:1957b,
	author = {Bardeen, J. and Cooper, L. N. and Schrieffer, J. R.},
	doi = {10.1103/PhysRev.108.1175},
	journal = {Phys. Rev.},
	number = {5},
	pages = {1175-1204},
	title = {Theory of Superconductivity},
	volume = {108},
	year = {1957}}

@article{bardeen:1957a,
	author = {Bardeen, J. and Cooper, L. N. and Schrieffer, J. R.},
	doi = {10.1103/PhysRev.106.162},
	journal = {Phys. Rev.},
	number = {1},
	pages = {162-164},
	title = {Microscopic Theory of Superconductivity},
	volume = {106},
	year = {1957}}

@article{boguslawski:2014a,
	author = {Boguslawski, K. and Tecmer, P. and Ayers, P. W. and Bultinck, P. and De Baerdemacker, S. and Van Neck, D.},
	doi = {10.1103/PhysRevB.89.201106},
	journal = {Phys. Rev. B},
	number = {20},
	pages = {201106(R)},
	title = {Efficient description of strongly correlated electrons with mean-field cost},
	volume = {89},
	year = {2014}}

@article{boguslawski:2014b,
	author = {Boguslawski, K. and Tecmer, P. and Bultinck, P. and De Baerdemacker, S. and Van Neck, D. and Ayers, P. W.},
	doi = {10.1021/ct500759q},
	journal = {J. Chem. Theory Comput.},
	number = {11},
	pages = {4873-4882},
	title = {Nonvariational Orbital Optimization Techniques for the {AP1roG} Wave Function},
	volume = {10},
	year = {2014}}

@article{boguslawski:2014c,
	author = {Boguslawski, K. and Tecmer, P. and Limacher, P. A. and Johnson, P. A. and Ayers, P. W. and Bultinck, P. and De Baerdemacker, S. and Van Neck, D.},
	doi = {10.1063/1.4880820},
	journal = {J. Chem. Phys.},
	number = {21},
	pages = {214114},
	title = {Projected seniority-two orbital optimization of the antisymmetric product of one-reference orbital geminal},
	volume = {140},
	year = {2014}}

@article{boguslawski:2015,
	author = {Boguslawski, K. and Ayers, P. W.},
	doi = {10.1021/acs.jctc.5b00776},
	journal = {J. Chem. Theory Comput.},
	number = {11},
	pages = {5252-5261},
	title = {Linearized Coupled Cluster Correction on the Antisymmetric Product of 1-Reference Orbital Geminals},
	volume = {11},
	year = {2015}}

@article{boguslawski:2016a,
	author = {Boguslawski, K. and Tecmer, P. and Legeza, \"{O}.},
	doi = {10.1103/PhysRevB.94.155126},
	journal = {Phys. Rev. B},
	number = {15},
	pages = {155126},
	title = {Analysis of two-orbital correlations in wave functions restricted to electron-pair states},
	volume = {94},
	year = {2016}}

@article{boguslawski:2017,
	author = {Boguslawski, K. and Tecmer, P.},
	doi = {10.1021/acs.jctc.6b01134},
	journal = {J. Chem. Theory Comput.},
	number = {12},
	pages = {5966-5983},
	title = {Benchmark of Dynamic Electron Correlation Models for Seniority-Zero Wave Functions and Their Application to Thermochemistry},
	volume = {13},
	year = {2017}}

@article{boguslawski:2021,
	author = {Boguslawski, K.},
	doi = {10.1039/D1CC04539C},
	journal = {Chem. Commun.},
	number = {92},
	pages = {12277-12280},
	title = {Open-shell extensions to closed-shell {pCCD}},
	volume = {57},
	year = {2021}}

@article{bytautas:2011,
	author = {Bytautas, L. and Henderson, T. M. and Jimenez-Hoyos, C. A. and Ellis, J. K. and Scuseria, G. E.},
	doi = {10.1063/1.3613706},
	journal = {J. Chem. Phys.},
	number = {4},
	pages = {044119},
	title = {Seniority and orbital symmetry as tools for establishing a full configuration interaction hierarchy},
	volume = {135},
	year = {2011}}

@article{bytautas:2015,
	author = {Bytautas, L. and Scuseria, G. E. and Ruedenberg, K.},
	doi = {10.1063/1.4929904},
	journal = {J. Chem. Phys.},
	number = {9},
	pages = {094105},
	title = {Seniority number description of potential energy surfaces: Symmetric dissociation of water, \text{N}$_2$, \text{C}$_2$, and \text{Be}$_2$},
	volume = {143},
	year = {2015}}

@article{coleman:1965,
	author = {Coleman, A. J.},
	doi = {0.1063/1.1704794},
	journal = {J. Math. Phys.},
	number = {9},
	pages = {1425-1431},
	title = {{Structure of fermion density matrices II. {A}ntisymmetrized geminal powers}},
	volume = {6},
	year = {1965}}

@article{cullen:1996,
	author = {Cullen, J.},
	doi = {10.1016/0301-0104(95)00321-5},
	journal = {Chem. Phys.},
	pages = {217-229},
	title = {Generalized valence bond solutions from a constrained coupled cluster method},
	volume = {202},
	year = {1996}}

@article{cullen:1999,
	author = {Cullen, J.},
	doi = {10.1002/(SICI)1096-987X(19990730)20:10%3C999::AID-JCC2%3E3.0.CO;2-A},
	journal = {J. Comput. Chem.},
	number = {10},
	pages = {999-1008},
	title = {{Is GVB-CI superior to CASSCF?}},
	volume = {20},
	year = {1999}}

@article{cullen:2007,
	author = {Cullen, J.},
	doi = {10.1002/jcc.20808},
	journal = {J. Comput. Chem.},
	number = {4},
	pages = {497-504},
	title = {An approximate diatomics in molecules formulation of generalized valence bond theory},
	volume = {29},
	year = {2008}}

@article{epstein:1926,
	author = {Epstein, P. S.},
	doi = {10.1103/PhysRev.28.695},
	journal = {Phys. Rev.},
	number = {4},
	pages = {695},
	title = {The {Stark} Effect from the Point of View of {Sc}hroedinger's Quantum Theory},
	volume = {28},
	year = {1926}}

@article{faribault:2008,
	author = {Faribault, A. and Calabrese, P. and Caux, J.-S.},
	title = {Exact mesoscopic correlation functions of the {R}ichardson pairing model},
	journal = {Phys. Rev. B},
	number = {6},
	pages = {064503},
	volume = {77},
	doi = {10.1103/PhysRevB.77.064503},
	year = {2008}}

@article{faribault:2010,
	author = {Faribault, A. and Calabrese, P. and Caux, J.-S.},
	title = {Dynamical correlation functions of the mesoscopic pairing model},
	journal = {Phys. Rev. B},
	number = {17},
	pages = {174507},
	volume = {81},
	doi = {10.1103/PhysRevB.81.174507},
	year = {2010}}

@article{fecteau:2022,
	author = {Fecteau, C.-\'{E}. and Cloutier, S. and Moisset, J.-D. and Boulay, J. and Bultinck, P. and Faribault, A. and Johnson, P. A.},
	doi = {10.1063/5.0091338},
	journal = {J. Chem. Phys.},
	number = {19},
	pages = {194103},
	title = {{Near--exact} treatment of seniority{--zero} ground and excited states with a {Richardson--Gaudin} mean{--field}},
	volume = {156},
	year = {2022}}

@article{gaudin:1976,
	author = {Gaudin, M.},
	doi = {10.1051/jphys:0197600370100108700},
	journal = {J. Phys. II},
	number = {10},
	pages = {1087-1098},
	title = {Diagonalisation d'une classe d'hamiltoniens de spin},
	volume = {37},
	year = {1976}}

@article{goddard:1967,
	author = {Goddard III, W. A.},
	doi = {10.1103/PhysRev.157.81},
	journal = {Phys. Rev.},
	number = {1},
	pages = {81},
	title = {Improved Quantum Theory of Many-Electron Systems. {II. T}he Basic Method},
	volume = {157},
	year = {1967}}

@article{goddard:1973,
	author = {Goddard III, W. A. and Dunning, T. H. and Hunt, W. J. and Hay, P. J.},
	doi = {10.1021/ar50071a002},
	journal = {Acc. Chem. Res.},
	number = {11},
	pages = {368-376},
	title = {Generalized valence bond description of bonding in low-lying states of molecules},
	volume = {6},
	year = {1973}}

@article{goddard:1978,
	author = {Goddard III, W. A. and Harding, L. B.},
	doi = {10.1146/annurev.pc.29.100178.002051},
	journal = {Annu. Rev. Phys. Chem.},
	number = {1},
	pages = {363-396},
	title = {The Description of Chemical Bonding From Ab Initio Calculations},
	volume = {29},
	year = {1978}}

@article{hay:1972,
	author = {Hay, P. J. and Hunt, W. J. and Goddard III, W. A.},
	doi = {10.1021/ja00779a002},
	journal = {J. Am. Chem. Soc.},
	number = {24},
	pages = {8293-8301},
	title = {Generalized valence bond description of simple alkanes, ethylene, and acetylene},
	volume = {94},
	year = {1972}}

@article{henderson:2014b,
	author = {Henderson, T. M. and Bulik, I. W. and Stein, T. and Scuseria, G. E.},
	doi = {10.1063/1.4904384},
	journal = {J. Chem. Phys.},
	number = {24},
	pages = {244104},
	title = {Seniority-based coupled cluster theory},
	volume = {141},
	year = {2014}}

@article{hunt:1972,
	author = {Hunt, W. J. and Hay, P. J. and Goddard III, W. A.},
	doi = {10.1063/1.1678308},
	journal = {J. Chem. Phys.},
	number = {2},
	pages = {738-748},
	title = {Self‐Consistent Procedures for Generalized Valence Bond Wavefunctions. {A}pplications {H}$_3$, {BH}, {H}$_2${O}, {C}$_2${H}$_6$, and {O}$_2$},
	volume = {57},
	year = {1972}}

@article{hurley:1953,
	author = {Hurley, A. C. and Lennard-Jones, J. E. and Pople, J. A.},
	doi = {10.1098/rspa.1953.0198},
	journal = {Proc. R. Soc.},
	number = {1143},
	pages = {446-455},
	title = {{The molecular orbital theory of chemical valency XVI. A theory of paired-electrons in polyatomic molecules}},
	volume = {A220},
	year = {1953}}

@article{johnson:2020,
	author = {Johnson, P. A. and Fecteau, C.-\'{E}. and Berthiaume, F. and Cloutier, S. and Carrier, L. and Gratton, M. and Bultinck, P. and De Baerdemacker, S. and Van Neck, D. and Limacher, P. and Ayers, P. W.},
	doi = {10.1063/5.0022189},
	journal = {J. Chem. Phys.},
	number = {10},
	pages = {104110},
	title = {{Richardson--Gaudin} mean-field for strong correlation in quantum chemistry},
	volume = {153},
	year = {2020}}

@article{johnson:2023,
	author = {Johnson, P. A. and DePrince III, A. E.},
	doi = {10.1021/acs.jctc.3c00807},
	journal = {J. Chem. Theory Comput.},
	number = {22},
	pages = {8129-8146},
	title = {Single reference treatment of strongly correlated {H$_4$} and {H$_{10}$} isomers with {Richardson--Gaudin} states},
	volume = {19},
	year = {2023}}

@article{johnson:2024b,
	author = {Johnson, P. A.},
	doi = {10.1021/acs.jpca.4c02857},
	journal = {J. Phys. Chem. A},
	pages = {6033-6045},
	title = {Beyond a {Richardson--Gaudin} Mean-Field: {Slater--Condon} Rules and Perturbation Theory},
	volume = {128},
	year = {2024}}

@article{kossoski:2021,
	author = {Kossoski, F. and Marie, A. and Scemama, A. and Caffarel, M. and Loos, P.-F.},
	doi = {10.1021/acs.jctc.1c00348},
	journal = {J. Chem. Theory Comput.},
	number = {8},
	pages = {4756-4768},
	title = {Excited States from State-Specific Orbital-Optimized Pair Coupled Cluster},
	volume = {17},
	year = {2021}}

@article{kossoski:2022,
	author = {Kossoski, F. and Damour, Y. and Loos, P.-F.},
	doi = {10.1021/acs.jpclett.2c00730},
	journal = {J. Phys. Chem. Lett.},
	number = {19},
	pages = {4342 - 4349},
	title = {Hierarchy Configuration Interaction: Combining Seniority Number and Excitation Degree},
	volume = {13},
	year = {2022}}

@article{kutzelnigg:1964,
	author = {Kutzelnigg, W.},
	doi = {10.1063/1.1725065},
	journal = {J. Chem. Phys.},
	number = {12},
	pages = {3640-3647},
	title = {Direct Determination of Natural Orbitals and Natural Expansion Coefficients of Many‐Electron Wavefunctions. {I. Natural} Orbitals in the Geminal Product Approximation},
	volume = {40},
	year = {1964}}

@article{kutzelnigg:2012,
	author = {Kutzelnigg, W.},
	doi = {10.1016/j.chemphys.2011.10.020},
	journal = {Chem. Phys.},
	pages = {119-124},
	title = {Separation of strong (bond-breaking) from weak (dynamical) correlation},
	volume = {401},
	year = {2012}}

@article{limacher:2013,
	author = {Limacher, P. A. and Ayers, P. W. and Johnson, P. A. and De Baerdemacker, S. and Van Neck, D. and Bultinck, P.},
	doi = {10.1021/ct300902c},
	journal = {J. Chem. Theory Comput.},
	number = {3},
	pages = {1394-1401},
	title = {A New Mean-Field Method Suitable for Strongly Correlated Electrons: Computationally Facile Antisymmetric Products of Nonorthogonal Geminals},
	volume = {9},
	year = {2013}}

@article{limacher:2014a,
	author = {Limacher, P. A. and Kim, T. D. and Ayers, P. W. and Johnson, P. A. and De Baerdemacker, S. and Van Neck, D. and Bultinck, P.},
	doi = {10.1080/00268976.2013.874600},
	journal = {Mol. Phys.},
	number = {5-6},
	pages = {853-862},
	title = {The influence of orbital rotation on the energy of closed-shell wavefunctions},
	volume = {112},
	year = {2014}}

@article{marie:2021,
	author = {Marie, A. and Kossoski, F. and Loos, P.-F.},
	journal = {J. Chem. Phys.},
	number = {10},
	pages = {104105},
	title = {Variational coupled cluster for ground and excited states},
	volume = {155},
	year = {2021}}

@article{moisset:2022a,
	author = {Moisset, J.-D. and Fecteau, C.-\'{E}. and Johnson, P. A.},
	doi = {10.1063/5.0088602},
	journal = {J. Chem. Phys.},
	number = {21},
	pages = {214110},
	title = {Density matrices of seniority{--zero} geminal wavefunctions},
	volume = {156},
	year = {2022}}

@article{nesbet:1955,
	author = {Nesbet, R. K.},
	doi = {10.1098/rspa.1955.0134},
	journal = {Proc. R. Soc. Lond. Ser. A},
	number = {1182},
	pages = {312-321},
	title = {Configuration interaction in orbital theories},
	volume = {230},
	year = {1955}}

@article{parkhill:2009,
	author = {Parkhill, J. A. and Lawler, K. and Head-Gordon, M.},
	doi = {10.1063/1.3086027},
	journal = {J. Chem. Phys.},
	pages = {084101},
	title = {The perfect quadruples model for electron correlation in a valence active space},
	volume = {130},
	year = {2009}}

@article{pernal:2013,
	author = {Pernal, K.},
	doi = {10.1016/j.comptc.2012.08.022},
	journal = {Comput. Theor. Chem.},
	pages = {127-129},
	title = {The equivalence of the {P}iris {N}atural {O}rbital {F}unctional 5 ({PNOF5}) and the antisymmetrized product of strongly orthogonal geminal theory},
	volume = {1003},
	year = {2013}}

@article{piris:2011,
	author = {Piris, M. and Lopez, X. and Ruip\'{e}rez, F. and Matxain, J. M. and Ugalde, J. M.},
	doi = {10.1063/1.3582792},
	journal = {J. Chem. Phys.},
	number = {16},
	pages = {164102},
	title = {A natural orbital functional for multiconfigurational states},
	volume = {134},
	year = {2011}}

@article{piris:2017,
	author = {Piris, M.},
	doi = {10.1103/PhysRevLett.119.063002},
	journal = {Phys. Rev. Lett.},
	number = {6},
	pages = {063002},
	title = {Global Method for Electron Correlation},
	volume = {119},
	year = {2017}}

@article{richardson:1963,
	author = {Richardson, R. W.},
	doi = {10.1016/0031-9163(63)90259-2},
	journal = {Phys. Lett.},
	number = {6},
	pages = {277-279},
	title = {A RESTRICTED CLASS OF EXACT EIGEN-STATES OF THE PAIRING-FORCE {H}AMILTONIAN},
	volume = {3},
	year = {1963}}

@article{richardson:1964,
	author = {Richardson, R. W. and Sherman, N.},
	doi = {10.1016/0029-5582(64)90687-X},
	journal = {Nucl. Phys.},
	pages = {221-238},
	title = {Exact eigenstates of the pairing-force {Hamiltonian}},
	volume = {52},
	year = {1964}}

@article{richardson:1965,
	author = {Richardson, R. W.},
	doi = {10.1063/1.1704367},
	journal = {J. Math. Phys.},
	number = {7},
	pages = {1034-1051},
	title = {Exact Eigenstates of the Pairing‐Force {H}amiltonian. {II}},
	volume = {6},
	year = {1965}}

@article{siegbahn:1981,
	author = {Siegbahn, P. E. M. and Alml\"{o}f, J. and Heiberg, A. and Roos, B.},
	doi = {10.1063/1.441359},
	journal = {J. Chem. Phys.},
	number = {4},
	pages = {2384-2396},
	title = {The complete active space {SCF (CASSCF)} method in a {Newton-Raphson} formulation with application to the {HNO} molecule},
	volume = {74},
	year = {1981}}

@article{silver:1969,
	author = {Silver, D. M.},
	doi = {10.1063/1.1671025},
	journal = {J. Chem. Phys.},
	number = {12},
	pages = {5108-5116},
	title = {Natural orbital expansion of interacting geminals},
	volume = {50},
	year = {1969}}

@article{small:2009,
	author = {Small, D. W. and Head-Gordon, M.},
	doi = {10.1063/1.3069296},
	journal = {J. Chem. Phys.},
	number = {8},
	pages = {084103},
	title = {Tractable spin-pure methods for bond breaking: Local many-electron spin-vector sets and an approximate valence bond model},
	volume = {130},
	year = {2009}}

@article{stein:2014,
	author = {Stein, T. and Henderson, T. M. and Scuseria, G. E.},
	journal = {J. Chem. Phys.},
	title = {Seniority zero pair coupled cluster doubles theory},
	doi = {10.1063/1.4880819},
	number = {21},
	pages = {214113},
	volume = {140},
	year = {2014}}

@article{tecmer:2014,
	author = {Tecmer, P. and Boguslawski, K. and Johnson, P. A. and Limacher, P. A. and Chan, M. and Verstraelen, T. and Ayers, P. W.},
	doi = {10.1021/jp502127v},
	journal = {J. Phys. Chem.},
	number = {39},
	pages = {9058-9068},
	title = {Assessing the Accuracy of New Geminal-Based Approaches},
	volume = {A118},
	year = {2014}}

@article{vanvoorhis:2000,
	author = {Van Voorhis, T. and Head-Gordon, M.},
	doi = {10.1063/1.481138},
	journal = {J. Chem. Phys.},
	number = {13},
	pages = {5633-5638},
	title = {A nonorthogonal approach to perfect pairing},
	volume = {112},
	year = {2000}}

@article{vanvoorhis:2000b,
	author = {Van Voorhis, T. and Head-Gordon, M.},
	doi = {10.1016/S0009-2614(99)01413-X},
	journal = {Chem. Phys. Lett.},
	number = {6},
	pages = {575-580},
	title = {The imperfect pairing approximation},
	volume = {317},
	year = {2000}}

@article{vanvoorhis:2001,
	author = {Van Voorhis, T. and Head-Gordon, M.},
	doi = {10.1063/1.1406536},
	journal = {J. Chem. Phys.},
	number = {17},
	pages = {7814-7821},
	title = {Connections between coupled cluster and generalized valence bond theories},
	volume = {115},
	year = {2001}}

@article{wang:2019,
	author = {Wang, Q. and Zou, J. and Xu, E. and Pulay, P. and Li S.},
	doi = {10.1021/acs.jctc.8b00854},
	journal = {J. Chem. Theory Comput.},
	number = {1},
	pages = {141-153},
	title = {Automatic Construction of the Initial Orbitals for Efficient Generalized Valence Bond Calculations of Large Systems},
	volume = {15},
	year = {2019}}

@article{wang:2022,
	author = {Wang, J. and Baerends, E. J.},
	title = {Self-Consistent-Field Method for Correlated Many-ELectron Systems with an Entropic Cumulant Energy},
	journal = {Phys. Rev. Lett.},
	number = {1},
	pages = {013001},
	volume = {128},
	doi = {10.1103/PhysRevLett.128.013001},
	year = {2022}}

@article{yang:1962,
	author = {Yang, C. N.},
	title = {Concept of Off-Diagonal Long-Range Order and the Quantum Phases of Liquid {H}e and of Superconductors},
	journal = {Rev. Mod. Phys.},
	number = {4},
	pages = {694-704},
	volume = {34},
	doi = {10.1103/RevModPhys.34.694},
	year = {1962}}

@article{zhou:2002,
	author = {Zhou, H.-Q. and Links, J. and McKenzie, R. H. and Gould, M. D.},
	title = {Superconducting correlations in metallic nanoparticles: {E}xact solution of the {BCS} model by the algebraic {B}ethe ansatz},
	journal = {Phys. Rev. B},
	number = {6},
	pages = {060502},
	volume = {65},
	doi = {10.1103/PhysRevB.65.060502},
	year = {2002}}

@article{zou:2020,
	author = {Zou, J. and Niu, K. and Ma, H. and Li, S. and Fang, W.},
	doi = {10.1021/acs.jpca.0c05216},
	journal = {J. Phys. Chem. A},
	number = {40},
	pages = {8321-8329},
	title = {Automatic Selection of Active Orbitals from Generalized Valence Bond Orbitals},
	volume = {124},
	year = {2020}}

@book{bartlett_book,
	address = {Cambridge},
	author = {Shavitt, I. and Bartlett, R. J.},
	publisher = {Cambridge University Press},
	subtitle = {MBPT and Coupled-Cluster Theory},
	title = {Many-Body Methods in Chemistry and Physics},
	year = {2009}}

@book{baxter_book,
	address = {New York},
	author = {Baxter, R.},
	publisher = {Dover},
	title = {Exactly Solved Models in Statistical Mechanics},
	year = {2007}}

@book{helgaker_book,
	address = {West Sussex},
	author = {Helgaker, T. and J{\o}rgensen, P. and Olsen, J.},
	publisher = {Wiley \& Sons},
	title = {Molecular Electronic-Structure Theory},
	year = {2000}}

@book{surjan_book,
	address = {Berlin},
	author = {Surj\'an, P. R.},
	publisher = {Springer},
	title = {An Introduction to the Theory of Geminals},
	year = {1999}}

@article{veillard:1967,
	author = {Veillard, A. and Clementi, E.},
	doi = {10.1007/BF01151915},
	journal = {Theor. Chem. Acc.},
	pages = {134-143},
	title = {Complete multi-configuration self-consistent field theory},
	volume = {7},
	year = {1967}}

\end{document}